\documentclass[12pt]{article}
\usepackage[margin=2.5cm]{geometry}  
\usepackage{epsfig, graphics}
\usepackage{longtable}
\usepackage{xcolor}
\usepackage{amsmath}
\usepackage[utf8]{inputenc} 
\usepackage[T1]{fontenc}
\usepackage{accents}
\usepackage{subcaption}
\usepackage{booktabs,caption}
\usepackage[flushleft]{threeparttable}
\usepackage{caption}
\usepackage{float}
\usepackage{bm}
\usepackage{graphicx}
\usepackage{amsmath}
\usepackage{natbib}
\usepackage{booktabs}
\usepackage{amsfonts}
\usepackage{amssymb}
\usepackage{hyperref}
\usepackage{ltxtable, filecontents}
\usepackage{tabularx}
\usepackage{setspace}
\usepackage{amsmath}
\newcommand{\nin}{\noindent}

\def\bkR{{\rm I\kern-.17em R}}

\def \id{1\hskip -3pt \mbox{l}}
\def \1n{1\hskip -3pt \mbox{N}}

\def \Frac {\displaystyle \frac }

\newfont{\bbf}{cmbx12 scaled 1435}

\newcommand{\resultpanel}[3]{%
  \begin{minipage}{0.47\linewidth}
    \centering
    \IfFileExists{#1}{%
      \includegraphics[width=\linewidth,height=0.32\textheight,keepaspectratio]{#1}
    }{%
      \fbox{\parbox[c][0.27\textheight][c]{0.9\linewidth}{\centering Missing figure\\[-1pt]\tiny\texttt{\detokenize{#1}}}}
    }

    \vspace{0.05cm}
    #2 #3
  \end{minipage}%
}

\begin{document}
\setlength{\baselineskip}{.26in}

\thispagestyle{empty}
\renewcommand{\thefootnote}{\fnsymbol{footnote}}
\vspace*{0cm}
\begin{center}

\setlength{\baselineskip}{.32in}
{\bf Recurrent Neural Networks for Forecasting Autoregressive Noncausal Processes  }\\
\setlength{\baselineskip}{.26in}
\vspace{0.4in}
Joann Jasiak\footnote[1]{York University, Canada, {\it e-mail}:
{\tt jasiakj@yorku.ca}},  Cheng Zhong\footnote[2]{Concordia University, Canada, {\it e-mail}:
{\tt c\_zhong2@live.concordia.ca} \\
The authors gratefully acknowledge financial support of the Natural Sciences and Engineering Research Council of Canada (NSERC: RGPIN-2024-06211 and RGPIN-2021-04144) and Social Sciences and Humanities Research Council (SSHRC: 430-2024-01325). The authors thank Christian Gourieroux and Yang Lu for helpful comments.}

\today\\

\medskip

\vspace{0.3in}
\begin{minipage}[t]{12cm}
\small
\begin{center}
Abstract \\
\end{center}
This paper examines the use of recurrent neural networks and their extensions for forecasting autoregressive causal-noncausal processes, characterized by 
local explosive patterns, including bubbles and spikes. We study recurrent neural networks (RNNs), long short-term memory networks (LSTMs), gated recurrent units (GRUs), bidirectional LSTMs, and also a simple feed-forward network (FNN) to reveal the difficulty of forecasting extreme values depending on the type of network and the length of the training set.
We show that FNN outperforms other networks in predicting extreme values of a univariate noncausal time series during a bubble. We also examine the improvements of point predictions of extreme values resulting from pooling and a modified loss function.  \\


{\bf JEL codes:} 
\end{minipage}

\end{center}
\renewcommand{\thefootnote}{\arabic{footnote}}

\newpage

\section{Introduction}

This paper examines the use of recurrent neural networks and their extensions to forecast univariate stationary autoregressive causal-noncausal autoregressive processes. These processes, referred to as Mixed Autoregressive (MAR(p,q)) are characterized by 
local explosive patterns such as bubbles and spikes. The size and frequency of local explosions depend on the distribution and serial dependence of a MAR. We study the recurrent neural networks (RNN), long short-term memory networks (LSTM), gated recurrent unit (GRU), Bidirectional LSTM and a simple feed-forward network (FNN) and compare their   
predictions of extreme values that occur during a bubble. We observe that generally, the recurrent neural networks have difficulty predicting extreme values, and their forecast accuracy depends on the training set, which needs to include explosive patterns to be predicted. To improve the accuracy of point forecasts, we explore pooling of training sets and propose to modify the loss function by replacing the mean squared prediction error by using an alternative non-MSE criteria, more adequate in the presence of bubbles. 

In the literature, there exist classical forecasting methods for univariate MAR processes. The simulation-based methods providing point predictions were introduced by \citet{LanneMarkku2012BMSA,LanneMarkku2012Ofon}. \citet{jasiak2016} and \citet{GourierouxChristian2021Fpab} introduced closed-form formulas of predictive density for univariate noncausal processes and explored their performance in application to MAR processes with heavy-tailed distributions. The predictive density formula in closed form was extended further to multivariate processes for predictive density forecasting and back-casting in \citet{GourierouxChristian2026NFaI}. A simple closed-form point forecast formula for MAR processes with stable distributed errors has been introduced by \citet{FriesZakoian2019}. Its advantage is that it is easy to apply in practice, and provides satisfactory results for a general class of MAR processes with heavy-tailed errors.

The evaluation of deep learning models for time series predictions has been given attention in the literature. \citet{chandra2021evaluation} review the existing literature on predictions of stationary univariate time series. They find that bidirectional and encoder-decoder LSTM networks provide the best performance accuracy. 
\citet{SakoMpindaRodrigues2022} worked on comparing the RNN, LSTM, and GRU networks for forecasting eight stock market indices and six exchange rates in both univariate and multivariate specifications. They argued that GRU has better forecasting performance in general, particularly for univariate out-of-sample exchange-rate forecasting and multivariate stock-index forecasting.
\citet{Berger2026} compares simple RNN, LSTM, and GRU with ARMA models for univariate daily return forecasting. Their study found that a single-layer RNN provides a competitive alternative to ARMA, while additional depth and more complex gated architectures introduced in LSTM and GRU, do not consistently improve forecasting accuracy.
\citet{ForoutanLahmiri2024} compare LSTM, GRU, their bidirectional variants, and several convolutional and hybrid architectures for univariate next-day forecasting of West Texas Intermediate (WTI) and Brent crude oil, gold, and silver prices. They find that GRU-based models generally outperform their LSTM counterparts. Temporal convolutional networks achieve the best results for WTI, Brent, and silver, whereas a bidirectional GRU with a 30-day input window performs best for gold. However, this literature does not consider predictions of bubbles and extreme values in a stationary time series.

The current interest in machine learning methodology raises the question of how accurate the deep learning methods are in application to forecasting bubbles and spikes. This paper addresses this question by providing theoretical results and simulation-based evidence. We illustrate the impact of the training set on the prediction of extreme values, and introduce improved predictive criteria that account for the extreme values that occur during a bubble. Moreover, we focus our attention on recursive one-step-ahead predictions, which is most commonly used in practice and easy to implement, based on PyTorch or Keras. A forecasting method based on a multilayer perceptron (MLP) with hidden layers for Mixed Autoregressive Moving Average (MARMA) models has been introduced recently by \citet{DumitrescuPeignonThomas2026}. That paper considers estimation of the predictive density of a process and compares the MLP with the aforementioned predictive density-based methods existing in the literature. Our contribution is to  point out the difficulties with  predicting extreme values from commonly used neural nets and to offer simple adjustments that can be easily implemented regardless of the dynamic model and error distribution.

The paper is organized as follows: Section 2 describes the univariate causal-noncausal MAR processes. Section 3 discusses the methodology underlying the Recurrent Neural Networks (RNN) and describes the RNN, FNN, LSTM, GRU and Bidirectional LSTM. Section 4 explores analytically the performance of various types of RNN in an application to heavy-tailed MAR processes with bubbles. Section 5 examines the performance of various types of RNN in simulations, shows the effects of pooling and introduces the LSTM with a modified loss function. Section 6 concludes.

\setcounter{equation}{0}\def\theequation{2.\arabic{equation}}
\section{Methodology}

The mixed noncausal autoregressive MAR(1,1) process is defined as the unique stationary solution of:
\begin{equation}
(1- \phi L)(1-\psi L^{-1}) y_t = \varepsilon_t, 
\end{equation}

\nin where the errors are non-Gaussian, independent, identically distributed and such that $E(|\varepsilon_t|^{\delta})<\infty$ for $\delta>0$. Parameters $\phi$ and $\psi$ are two
autoregressive coefficients that are strictly less than one. 
Coefficient $\phi$ represents the standard causal persistence while coefficient $\psi$ depicts noncausal persistence.

When the distribution of $\varepsilon_t$ has fat tails, the parameters acquire an additional interpretation, as a large negative error value creates a spike in the trajectory with an explosion rate of about $1/\psi$ and a collapse rate of $\phi$. We observe a jump if $\psi = 0$ and $\phi>0$ and an explosive bubble if $\psi$ is small, positive and $\phi=0$. 
When $\psi= 0$, equation (2.1) defines a purely causal autoregressive process. If $\phi= 0$, the equation above defines a purely noncausal process. If both polynomials contain non-zero coefficients, then equation (2.1) describes a mixed causal-noncausal MAR(1,1) process. The mixed process contains both leads and lags of $y_t$, and admits a two-sided moving average representation [Gourieroux, Zakoian (2015)]. 

It follows from Lanne, Saikkonen (2011),  and Lanne, Luoto, Saikkonen
(2012),  that process ($y_t$) has the following unobserved components $u_t, v_t$ defined by:
\begin{equation}
  u_t (1- \phi L)  y_t \leftrightarrow (1-\psi L^{-1})  u_t = \varepsilon_t, \;\;\;\;
  v_t \equiv  (1-\psi L^{-1})  y_t \leftrightarrow (1- \phi L)   v_t = \varepsilon_t,
\end{equation}

\nin which can be interpreted as the "causal" and "noncausal"
components of process ($y_t$), respectively. Gourieroux, Jasiak (2016) show that  i) $u_t$ is $\varepsilon$-noncausal
and $y$-causal and ii) $v_t$ is $\varepsilon$-causal and
$y$-noncausal. Process $y_t$  has the following representations based on unobserved components:
 
\begin{equation}
y_t = \Frac{1}{1-\phi \psi} (\phi v_{t-1} + u_t), \;\;
y_t = \Frac{1}{1-\phi \psi} (v_t + \psi u_{t+1}) .
\end{equation}

\noindent In the above representations, $y_t$ is either a linear function of the
current value of the $\varepsilon$-causal component $v$ and of the 
first lag of the $\varepsilon$-noncausal component $u$, or vice-versa. 

\subsection{Filtering and Simulation}

Gourieroux, Jasiak (2016) show that the deterministic relation (2.2) allows us to filter the unobserved components given the observations on process $(y_t)$, over a period of length $T$. Let 
$(y_1, \ldots, y_T)$ denote the observed sequence.

\nin{\bf Filtering Algorithm}

The values of unobserved
components $u$ and $v$ and errors $\varepsilon$ can be computed
from a set of observations $(y_1, \ldots, y_T)$ as follows:

\noindent (i) From equation (2.1) for $t =2, \ldots, T-1$, we obtain the
values $\varepsilon_{2}, \ldots, \varepsilon_{T-1}$ as functions
of $(y_1, \ldots, y_T)$.\vspace{1em}

\noindent (ii) From equation (2.2)~: $u_t = (1- \phi L)   y_t, t = 2,
\ldots,T$, we obtain $u_{2},\ldots, u_T$.\vspace{1em}

\noindent (iii) From equation (2.3) : $v_t = (1-\psi L^{-1})   y_t,
t=1,\ldots,T-1$, we obtain $v_1, \ldots, v_{T-1}$.\vspace{1em}

\nin When an additional observation $y_{T+1}$ becomes available, 
the set of unobserved components can be updated by computing 
$\varepsilon_{T}, u_{T+1}$ and $v_{T}$.

This algorithm is valid for any pair of parameters $\phi, \psi$. In particular, it is valid for the (unknown) true values $\phi_0, \psi_0$ and the true errors and unobserved components. It is also valid for other values of $\phi, \psi$. In that case, we get residuals $\hat{\varepsilon}_t$ and components $\hat{u}_t, \hat{v}_t$ that depend on $\phi, \psi$. Among these, $\hat{u}_T$ has an important role, as shown later in the text.

\medskip
\nin {\bf Simulation Algorithm (\citet{jasiak2016})}

Formulas (2.2)- (2.3) can also be used to simulate the trajectories of process $(y_t)$ from the simulated unobserved components as follows:

\medskip

\noindent step 1: Simulate a path of i.i.d. errors $\varepsilon_t^s, \; t=1,...,T$.

\medskip

\noindent step 2: Use formula (2.2) to obtain the simulated paths of the $\varepsilon$-causal 
and $\varepsilon$-noncausal components :

$$u_t^s = \varepsilon_t^s + \psi u_{t+1}^s,\; t =1 ,...,2T,$$ 

$$v_t^s = \varepsilon_t^s + \phi v_{t-1}^s,\; t =-T ,...,T,$$

\noindent starting from a far terminal condition (resp. far initial condition) $u_{2T}^s=u_0$, say
(resp. $v_{-T}^s=v_0$).

\medskip

\noindent step 3: The simulated trajectory $(y_t^s)$ is obtained from
either one of the two partial fraction
representations in (2.3).

\subsection{Forecasting}

For forecasting the MAR processes,  Gourieroux, Jasiak (2016) introduce the predictive density estimator.
The predictive density has a closed-form plug-in expression when the errors are Cauchy distributed, or more generally, follow an $\alpha$-stable distribution. For example, when $\varepsilon_t$ follows a Cauchy distribution, 
the one-step ahead predictive density of MAR(1,1) is given by:
$$
\pi(u_{T+1}|u_T) = \frac{1}{\pi} \frac{1}{1+(u_T - \psi u_{T+1})^2} 
\frac{1+(1-\psi)^2 u_T^2}{1+(1-\psi)^2 u_{T+1}^2 },
$$

\nin The estimator of the predictive density can be used to forecast $y_{T+1}$ as follows. Given that
$y_{T+1} = u_{T+1}^s + \phi y_{T},$ one can compute the predictive  density $\hat{\pi}(y_{T+1}|y_T, \hat{u}_T)$ by shifting the predictive density $\hat{\pi} (u_{T+1}| \hat{u}_{T})$ of $u_{T+1}$ by $\phi y_T$, when $\phi $ is known, or by $\hat{\phi} y_T$ when $\phi $ is estimated. 

The formula of predictive density of a MAR process can be computed either semi-parametrically or parametrically following the approach in Gourieroux, Jasiak (2016), or Gourieroux, Jasiak (2025) for univariate and multivariate causal-noncausal processes, respectively. Then, the (top) mode of the predictive density (Gourieroux, Jasiak (2016)) can be used for point forecasting and the tail quantiles can be used to build forecast intervals at a desired level.

When the error distribution of a univariate MAR process is known to be $\alpha$-stable, with not too heavy tails, i.e. $\alpha \in (0.5, 2)$, then the point forecast can be computed analytically from the past values of the process, using its semi-strong representation, as shown in \citet{FriesZakoian2019}. 

The one-step ahead forecast formula is particularly simple for a MAR(1,1) with Cauchy distributed errors ($\alpha=1$) and a positive coefficient $\psi$. It is given by:

\begin{equation}
E(Y_{t+1}| {\cal F}_t) = Y_t + (1-L) \phi Y_t = Y_t + \phi (Y_t -Y_{t-1}),
\end{equation}

\nin where ${\cal F}_t$ denotes the canonical filtration of process $(Y_t)$ [see \citet{FriesZakoian2019}, Proposition 3.2.]. This simple forecast formula is used later in Section 5.

\setcounter{equation}{0}\def\theequation{3.\arabic{equation}}
\section{RNN Methodology}

This section describes various types of neural networks used for point predictions in time series analysis. Below, we examine the RNN, LSTM, FNN, GRU and Bidirectional LSTM.

\subsection{Recurrent Neural Networks}

\begin{figure}[H]
    \centering
    \caption{Recurrent Neural Network Architecture}

    \begin{subfigure}[b]{\textwidth}
        \centering
        \includegraphics[height=5cm]{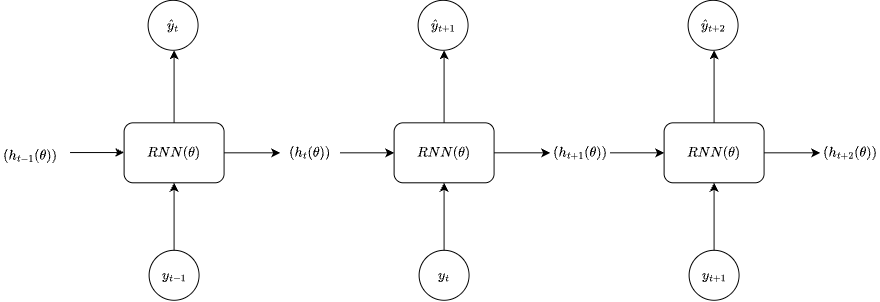}
        \caption{RNN Architecture with Respect to Time}
        \label{rnntime}
    \end{subfigure}

    \vspace{0.5cm}

    \begin{subfigure}[b]{\textwidth}
        \centering
        \includegraphics[height=7cm]{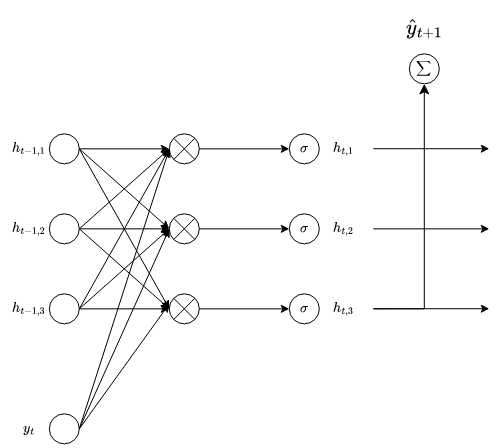}
        \caption{The Internal Architecture of RNN Unit}
        \label{rnnact}
    \end{subfigure}
\end{figure}

Recurrent Neural Networks (RNNs) and their extensions, including Long Short-Term Memory (LSTM) and Gated Recurrent Unit (GRU) models, have been widely applied in sequence modeling. Due to the sequential nature of time-series data, RNN-based approaches are widely used and particularly suited for forecasting \citep{LimZohren2021}. RNNs keep input information from the previous step and pass it to the next input. It contains a context layer that maintains the memory of the prior hidden layer. In the simplest univariate case, the one-step-ahead forecast of a time series $(y_t)$ can be written as:

\begin{equation}
o_t = \hat{y}_{t+1} = W_y h_t + b_y,
\end{equation}

\nin where $y_t$ is the input at time $t$ fed into the hidden state vector $h_t$ written as:

\begin{equation}
    h_t = \sigma (W_{h,1} h_{t-1}+W_{h,2} y_t + b_h),
\end{equation}

\nin where $W_{.}$ and $b_{.}$ are weights and biases of the model, respectively, and $\sigma$ is the  activation function, such as the logistic or tanh function with outputs in $(0,1)$ and $(-1,1)$, respectively\footnote{Note that when $\hat{y}_{t+1} = \sigma(W_y h_t + b_y)$, the time series needs to be scaled \citep{chandra2021evaluation}}.

\subsection{Long Short-Term Memory}

LSTM networks were introduced to overcome the limitations of conventional RNNs by introducing more stable gradient propagation during training \citep{HochreiterSepp1997LSM}. 

\begin{figure}[H]
    \centering
    \caption{Long Short-Term Memory Architecture}

    \begin{subfigure}[b]{0.8\textwidth}
        \centering
        \includegraphics[height=4cm]{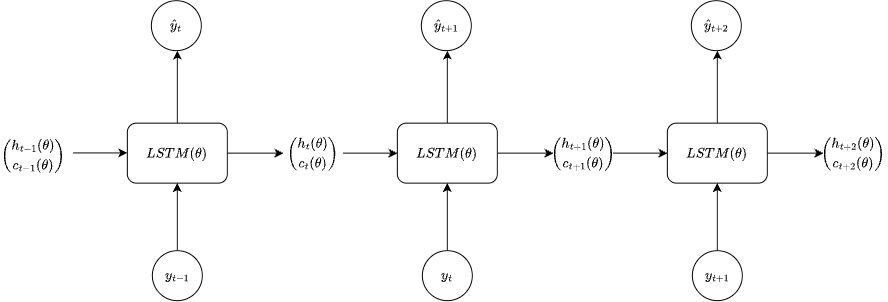}
        \caption{LSTM Architecture with Respect to Time}
        \label{lstmtime}
    \end{subfigure}
    \hfill
    \begin{subfigure}[b]{0.8\textwidth}
        \centering
        \includegraphics[height=4cm]{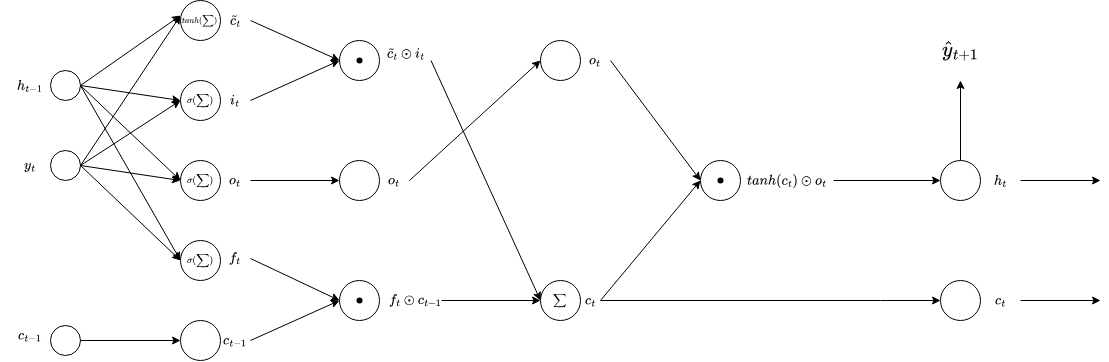}
        \caption{The Internal Architecture of LSTM Unit }
        \label{lstmact}
    \end{subfigure}
\end{figure}


A cell state $c_t$ is introduced to store long-term memory through gate mechanisms \citep{LimZohren2021}. The system of equations for an LSTM can be written as follows :

\begin{equation}
    f_t = \sigma(W_{f_1}\cdot y_t + W_{f_2}\cdot h_{t-1} + b_f)
\end{equation}

\begin{equation}
    i_t = \sigma(W_{i_1}\cdot y_t + W_{i_2} \cdot h_{t-1} + b_i)
\end{equation}

\begin{equation}
    \tilde{c}_t = tanh(W_{c_1}\cdot y_t + W_{c_2}\cdot h_{t-1} + b_c)
\end{equation}

\begin{equation}
    c_t = c_{t-1}\odot f_t + \tilde{c}_t \odot i_t
\end{equation}

\begin{equation}
    o_t = \sigma(W_{o_1} \cdot y_t+ W_{o_2}h_{t-1} + b_o)
\end{equation}

\begin{equation}
    h_t = tanh(c_t)\odot o_t
\end{equation}

\nin The final output of the one-step-ahead forecast $y_{t+1}$ is:

\begin{equation}
    \hat{y}_{t+1} = W_y\cdot h_{t} + b_y
\end{equation}

\nin where for  a univariate output (forecast) we have 
$
\hat y_{t+1}\in \mathbb{R}^{1\times 1}$,
$W_y\in \mathbb{R}^{1\times u}$,$b_y\in \mathbb{R}^{1\times 1}$.
For a multivariate (forecast) output of dimension $\ell$, we have:
$
\hat y_{t+1}\in \mathbb{R}^{\ell\times 1},
W_y\in \mathbb{R}^{\ell\times u},
b_y\in \mathbb{R}^{\ell\times 1},
$ where $u$ denotes the number of neurons $h_{t}$, i.e. the hidden states of the LSTM; $\sigma(.)$ and $tanh(.)$ represent the element-wise sigmoid activation function and the element-wise hyperbolic tangent activation function, respectively; and $\odot$ is the element-wise (Hadamard) product. The forget gate $f_t$, input gate $i_t$, and output gate $o_t$ are all determined by the previous hidden state $h_{t-1}$ and the current input $y_t$. 

In particular, the forget gate, $f_t$, determines the fraction of information from the previous time step that is removed from the memory state, while the input gate $i_t$ controls how much of the information from the candidate state $\tilde{c}_t$, determined from $h_{t-1}$ and  $y_t$,  is incorporated into the updated state $c_t$. The current hidden state $h_t$ is then obtained from the updated state $c_t$ through the output gate $o_t$. Finally, the LSTM cell outputs both $c_t$ and $h_t$, which are passed to the next time step. 

\[
\begin{aligned}
\hat y_{t+1}
=
W_y
\Bigg\{
\tanh
\Big[
&c_{t-1}\odot
\sigma\left(W_{f_1}y_t+W_{f_2}h_{t-1}+b_f\right) \\
&+
\tanh\left(W_{c_1}y_t+W_{c_2}h_{t-1}+b_c\right)
\odot
\sigma\left(W_{i_1}y_t+W_{i_2}h_{t-1}+b_i\right)
\Big] \\
&\odot
\sigma\left(W_{o1}y_t+W_{o2}h_{t-1}+b_o\right)
\Bigg\}
+
b_y
\end{aligned}
\]

\subsection{Gated Recurrent Unit}

Gated Recurrent Units (GRUs) have been shown in many applications to match or even outperform LSTMs, with the additional advantage of easier training \citep{NiuZhaoyang2023RauA}.

\begin{figure}[H]
    \centering
    \caption{Gated Recurrent Unit}

    \begin{subfigure}[b]{0.8\textwidth}
        \centering
        \includegraphics[width=\linewidth]{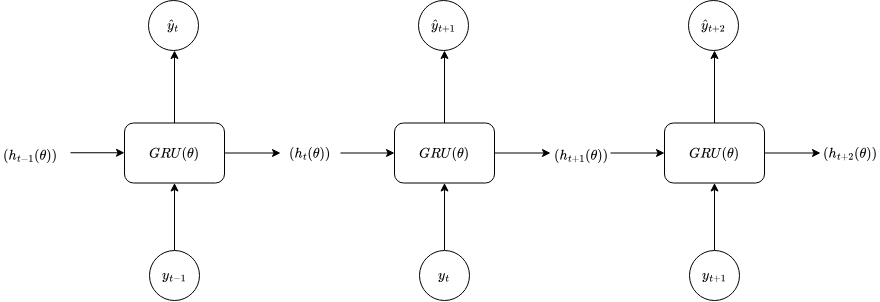}
        \caption{GRU Architecture with Respect to Time}
        \label{fig:gru}
    \end{subfigure}

    \vspace{0.5cm}

    \begin{subfigure}[b]{0.9\textwidth}
        \centering
        \includegraphics[width=\linewidth]{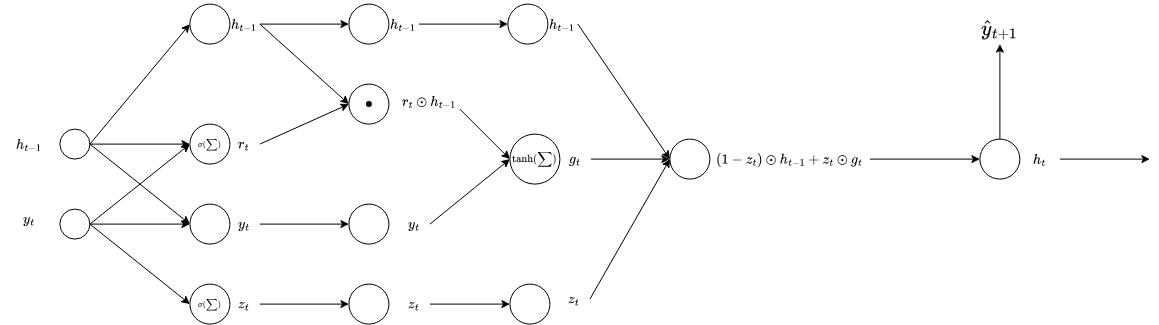}
        \caption{The Internal Architecture of a GRU Unit}
        \label{gru_cell}
    \end{subfigure}
\end{figure}

A single update gate was introduced by \citet{gru} to replace the forget and input gates, reducing the number of parameters required for training.

\begin{equation}
    r_t = \sigma(W_{r_1}\cdot y_t + W_{r_2}\cdot h_{t-1}+ b_r)
\end{equation}

\begin{equation}
    z_t = \sigma(W_{z_1}\cdot y_t + W_{z_2}\cdot h_{t-1} + b_z)
\end{equation}

\begin{equation}
    g_t = tanh(W_{g_2}\cdot y_t+ W_{g_2} \cdot (r_t\odot h_{t-1})+ b_g)
\end{equation}

\begin{equation}
    h_t = (1-z_t) \odot h_{t-1} + z_t\odot g_t
\end{equation}
\nin and
$$\hat{y}_{t+1} = W_y h_t + b_y$$

\nin where $r_t$ is the reset gate, which determines how the previous hidden state $h_{t-1}$ is used in constructing the candidate value $g_t$ together with the current input $y_t$. The update gate $z_t$ controls the proportion of the previous hidden state $h_{t-1}$ that is retained in the output hidden state $h_t$, while the remaining proportion, $ (1-z_t) $, is assigned to the candidate value. Under the standard GRU, the sigmoid activation function is applied, and hence $z_t \in[0,1]$. When $z_t=1$, the update gate is closed, and no memory from the previous hidden state is passed to the output hidden state, which implies full replacement by the candidate value. 

\subsection{Feed-Forward Neural Network}
To facilitate comparison with the RNN-based models, we also employ a commonly used two-layer feed-forward neural network (FNN) as a benchmark.

\begin{figure}[H]
    \centering
        \caption{FNN}
    \includegraphics[width=0.5\linewidth]{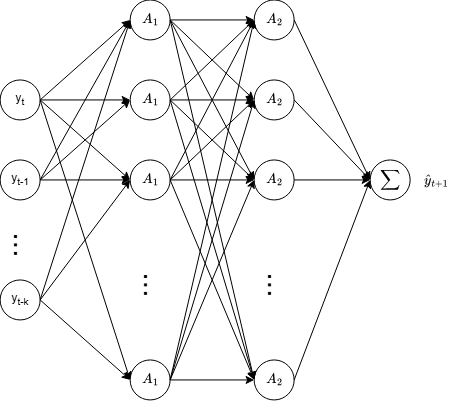}

    \label{fig:nn_figure}
\end{figure}

\nin The one-step-ahead forecast is:

$$
\hat{y}_{t+1} =  \sum_{k=1}^Kw_{2k} \, \sigma(\sum_{j=1}^J w_{1j} A_j +b_1) +b_2,
$$

\nin where

$$
A = \sigma(\sum_{i=0}^j w_{a,i} y_{t-i} +b_A)
$$

\nin $J$ and $K$ are the number of neurons used in each hidden layer. The activation function $\sigma$ is $relu$ by default in PyTorch with output in $(0, \infty)$ and the parameters $b_0, b_1$ and $b_2$ are bias terms in each layer. Hence, FNN involves less rescaling of the data by activation functions with outputs in $(0,1)$, or $(-1,1)$.

\subsection{Bidirectional LSTM}

\begin{figure}[H]
    \centering
        \caption{Bidirectional LSTM }
    \includegraphics[width=0.45\linewidth]{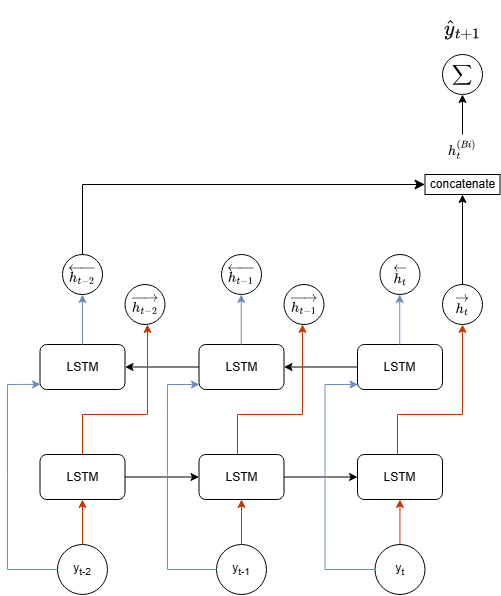}

    \label{fig:bilstm_figure}
\end{figure}

Bidirectional LSTM (BiLSTM) is designed to learn rules of both forward-looking and backward-looking by implementing two LSTM cells in two directions in the same layer. BiLSTM would intake hidden inputs in two ways, one from the past to future and one from the future to past, which differs from conventional LSTM networks. The model is shown in Figure \ref{fig:bilstm_figure}. 

The output of BiLSTM $h_t^{(Bi)}$ can be written as a combination of backward hidden state $\overleftarrow{h_t}$ and forward hidden state  $\overrightarrow{h_t}$. Below, we illustrate an example with the lookback window of length equal to $l$:

\begin{equation}
    \overrightarrow{h_t} = f(\overrightarrow{W_1} \cdot y_t + \overrightarrow{W_2} \cdot \overrightarrow{h}_{t-1} + \overrightarrow{b})
\end{equation}

\begin{equation}
    \overleftarrow{h}_{t-l+1} = f(\overleftarrow{W_1} \cdot y_{t-l+1} + \overleftarrow{W_2} \cdot \overleftarrow{h}_{t-l+2} + \overleftarrow{b})
\end{equation}

\begin{equation}
    h_t^{(\mathrm{Bi})}
=
\begin{bmatrix}
\overrightarrow{h_t}\\
\overleftarrow{h}_{t-l+1}
\end{bmatrix}
\end{equation}

\nin 

\begin{equation}
\hat{y}_{t+1}
=
g\left(W_y h_t^{(\mathrm{Bi})}+b_y\right)
\end{equation}

\nin where $h_t^{(\mathrm{Bi})}$ is usually the concatenation of $\overrightarrow{h_t}$ and $\overleftarrow{h}_{t-l+1}$.


This approach assumes that the process is time revertible with identical weights and biases of backward and forward equations which is not the case on non-Gaussian processes. It is more suitable for filtering purposes and for finding the missing values of a time series, given the values before and after. It is especially useful in the encoder-decoder application to sequences of words. In time series analysis, it is of limited use for out-of-sample forecasting $y_{T+1}$, as a future out-of-sample $h_{t+1}$ needs to be determined from a sufficiently large number of future observations, which are not available in practice.


\subsection{Tuning Parameters and NN-based Predicting}

We use an example of the LSTM prediction model to describe the NN-based prediction and explain the role of basic tuning parameters.

The  LSTM accepts input data structured as a 3D Tensor of dimension block formatted by (Batch Size, Time Steps/Sequence Length, i.e. window length, Feature Size).
For a sliding window look-back prediction model with a  window length of 10, for example, sequential training windows contain 10 consecutive values used to predict the next value of a time series.

An epoch in machine learning is one complete pass of the entire training dataset through a model. During a single epoch, every observation in the training data is processed, allowing the algorithm to update its internal weights and learn patterns.

During each epoch, all (disjoint or overlapping) windows of a given length
are passed through LSTM. The internal gates (forget, input, output) calculate state transitions, compute prediction loss and update weights in batches of a given multiple of windows.
On the final pass, the model processes the windows one last time with optimized weights, yielding the lowest training error.

By default, in standard sequential LSTM training, the windows appear in the exact same chronological order in each epoch. The alternative of shuffled windows, i.e. of a random window order needs to be avoided in time series analysis. Under that scenario, the LSTM internal memory states must be reset to zero between every window.

The window size needs to be between 1 and $T-1$, where $T$ is the length of a time series, because in order to train, LSTM needs at least one window plus one target observation to predict. LSTM applied to a  given window of length 10 (also called the time step), predicts the value of the first observation following chronologically that window.  

The parameters, i.e. weights and biases are re-estimated after each batch. A batch may contain several windows. Each of those windows generates a prediction. By comparing these predictions with the true observations, the prediction errors are calculated. Then, the optimisation algorithms, such as Adam, or SGD compute the gradients and re-estimate and update the weights and biases to reduce the mean square error. When batch size is equal to 1, the weights and biases are re-estimated after each window. A standard method in deep learning, applies LSTM to $N=4$ windows, accumulates the errors from these four windows, and re-estimates the parameters after that group of 4 windows. There also exists a mode "Batch Complete", where the batch size is equal to the total number of windows in the time series. Then, the weights and biases are re-estimated only once after the end of each epoch. For example, with 2 {\it  disjoint} windows of length 10,  and batch size of 1, after 10 epochs, the parameters will be re-estimated 20 times \footnote{A different number of re-estimations is obtained when overlapping windows are used.} .

The learning rate (LR) is an important parameter determining weight updating. While the gradient ($\Delta$) gives the direction of change for each parameter to reduce the mean square error (i.e. the loss function), the learning rate decides what proportion of that correction is applied to obtain a new weight: $w^p = w^{p-1}+ LR \times \Delta$  at optimisation step $p$.

 The neural networks are typically trained by the stochastic gradient descent  (SGD) algorithm. SGD maintains a single learning rate for all the parameter updates that does not change during training (unless specified and set manually). The Adam optimiser extends SGD by adapting the learning rate for each parameter (weight) as learning unfolds using the first and second moments of the gradients. The new optimizers, like Adam use an adaptive learning rate. Then, Adam reduces the LR for the weights that change too much and increases it for the parameters that move very little. 

An iteration is a single update step of the model's weights after processing a single batch. The total iterations in one epoch equal the total number of windows divided by the batch size. 

Models usually need many epochs (dozens, hundreds, or thousands) to reduce error and learn complex features.
The managing of epochs is needed to determine their optimal number to be used. Underfitting arises when too few epochs are used. It means the model does not finish learning the patterns, leading to low accuracy. Overfitting occurs when too many epochs are used that cause the model to memorize the training data too closely, making it perform poorly on new data.
Early stopping is a method that monitors the model's progress on a validation dataset and stops training automatically when improvement stops. 

Sometimes data is shuffled at the start of each epoch to prevent the model from memorizing the sequence of data.
An LSTM shuffles entire sequences (or batches of sequences) between each epoch, but it never shuffles the data inside individual windows. This ensures the internal chronological order of time steps remains intact while preventing the model from memorizing the order of the sequences themselves. The shuffling behaviour depends on whether one uses a Stateless LSTM or a Stateful LSTM. 

\nin 1. Stateless LSTMs

In a stateless LSTM, the cell states and hidden states are reset to zero at the beginning of every window or batch.
The data framework (like PyTorch DataLoader or Keras model.fit(shuffle=True)) randomly reorganizes the windows (also called blocks) before each epoch.
Because the LSTM forgets everything between separate sequences, it does not matter which sequence it reads first. Shuffling them reduces variance and prevents overfitting. 

For example:
Original Data:   [Sequence 1: A→B→C] , [Sequence 2: D→E→F] , [Sequence 3: G→H→I]

Epoch 1 Shuffle: [Sequence 2: D→E→F] , [Sequence 1: A→B→C] , [Sequence 3: G→H→I]

Epoch 2 Shuffle: [Sequence 3: G→H→I] , [Sequence 2: D→E→F] , [Sequence 1: A→B→C]

\medskip

\nin 2. Stateful LSTMs

In a stateful LSTM, the network carries its memory states over from the end of one window to the beginning of the next. This approach is used when a single continuous time series is too long and must be split into continuous chunks. 
We need to explicitly disable shuffling (shuffle=False).
If you shuffle the data, a chunk from next week might be fed right before a chunk from last month. This breaks the logical continuous timeline and destroys the network's ability to learn long-term dependencies. 

At the start of epoch two, three, etc, the parameters are set equal to those that minimize the loss function from the previous epoch, and the states are reset to zero, unless a trainable state option is used. 

Suppose, for example, that we have a sample of observations on a 1st-order autoregressive process (AR(1)) of length 300. For that AR(1) with 300 observations, one should choose between 50 and 100 epochs when estimating parameters by deep learning or gradient descent. Because an \(AR(1)\) model is mathematically simple and has very few parameters, it converges rapidly.
The count of 50 to 100 epochs is optimal because of
the low parameter count. An AR(1) model only estimates a constant (\(\mu \)), one lag coefficient (\(\phi _{1}\)), and noise variance (\(\sigma ^{2}\)). Moreover, there is a small dataset risk: With only 300 observations, running too many epochs will quickly cause overfitting, where the model memorizes the specific noise in your 300 points rather than learning the true underlying process.

For this example, it is recommended to use a batch size of 32 or 64 observations. For 300 observations, a batch size of 32 means that we will have \(\lceil 300 / 32 \rceil = 10\) iterations (weight updates) per epoch. 
To apply early stopping, one should not rely on a fixed epoch number alone. Instead, it is recommended to monitor a validation set (e.g., the last 60 observations) and stop training if the validation loss stops decreasing for 10 consecutive epochs (patience = 10). Moreover, it is useful to use a learning rate decay, i.e. to start with a standard learning rate (like \(0.001\) with the Adam optimizer) so the model settles smoothly into the global minimum within your 50–100 epoch window.

\setcounter{equation}{0}\def\theequation{4.\arabic{equation}}
\section{Application to MAR Processes}

In this section, we apply the RNN and their extensions to predict a MAR(1,1) process. There are several issues that arise in the application of RNN and their extensions to MAR(1,1) processes.

\subsection{The RNN Model}

1. {\it Hidden States}

As shown in Section 2, the MAR(1,1) processes rely on 2 hidden state variables. Both are stationary and can be written either as a causal or noncausal AR(1) in $y_t$. Equivalently, $y_t$ is a moving average of order infinity in the latent states.

In a basic neural net with $y_{t-1}$ as input, the prediction is $\hat{y}_{t+1} = w_y h_t + b_y = w_y \sigma( w_{h,1} h_{t-1} + w_{h,2} y_t + b_h) + b_y)$, which implies that $y_t$ has ARMA(1)-type of dynamics with hidden state $h_t$ instead of error term. Despite that $h_t$ plays the role of an "error" in this ARMA-type of representation, it is deterministic in the sense that all transitions are deterministic and the only source of randomness is $y_t$ itself. Hence, the hidden state $h_t$ can be written as an autoregressive process of order infinity in $y_t$ (AR($\infty$)), which explains its interpretation of "long memory" in the LSTM model, which is in fact a "long-range" memory rather than "long memory" in the sense of a hyperbolic decay rate of autocorrelations.  If several $h_t$'s are used in parallel, then the outcome has a more complicated nonlinear ARMA dynamics and the coefficients of a causal autoregressive representation of order infinity of a vector-valued $h_t$ in past $y_t$'s  are not necessarily summable in either mean square or absolute values, which may cause convergence problems. The neural nets applied to  a MAR(1,1) necessarily misfit the true hidden states and unnecessarily trace long autoregressive patterns from the past.

It follows that a large number of parallel hidden states is needed to approximate the ARMA($\infty$) representation of the causal state $u_t$. The increased number of parameters requires longer time series for identification. As mentioned earlier, the coefficients of such causal autoregressive of order infinity representation in $y_t$ are not necessarily summable in either mean square or absolute values, which may cause convergence problems.

\medskip
2. {\it Units}.

A unit in the LSTM layer is the hidden state, i.e. the number of hidden neurons inside the LSTM cell. It represents the memory size. Contrary to a classical neural net where 100 units means 100 independent neurons, in LSTM each unit is an element of the memory vector. A common choice of 128 units  means that $h_t$ and $c_t$ are each of size 128, and the gates $f_t, i_t , o_t$, the forget, input and output gates would each produce 128 values. A small number of units, like, 16, 32 is used for short memory, 64 or 128 for financial time series and 256, 512 for complex, nonlinear structures. The initial memory states, $h_0$ and $c_0$, are usually initialized to zero and treated as trainable parameters, but they may instead be held fixed at chosen values.

\medskip
3. {\it Parameter size}.

The choice of units determines the parameter size as follows. Let us denote the number of units by $dim(h)=dim(c) =m$ and the size of the time series by $dim(n)$. Then in each epoch, we have:
$$dim(\theta) = 4m(n+m+1)$$

\nin as the number of parameters. In the output equation, we have additionally
$nm +n$ parameters. Note that in the PyTorch software, the above formula is slightly different as it adds a bias per unit in each of 4 gates, which makes 8 additional parameters for $m=1$.

\nin In practice, if the initial values $h_0,c_0$ are trainable, the parameter number increases by $2m$. The numbers of parameters tend to be large, for example with $m=32$, we get 4096 parameters per epoch.

\medskip
4. {\it Vanishing Gradient}

The occurrence of extremely large positive or negative values creates estimation problems. The back-propagation-based gradients of SGD and Adam algorithms  accumulate over time
and layers. This procedure, which simplifies the calculations numerically,
is disadvantageous in the application to MAR(1,1) processes, as clusters of extreme values lead to zero gradients (vanishing gradient) by analogy to the difficulty encountered in fitting time series with a global trend.
Hence, the algorithm has a difficulty learning from past patterns.

Moreover, the RNN and their extensions rely heavily on averaging for bias reduction. For processes with infinite marginal mean and variance, the averages are not consistent estimators of the theoretical moments, which adds to the difficulty of estimation.

\medskip

5. {\it Loss Function}

The MSE is a poor loss function for variables with possibly infinite moments, such as MAR(1,1) processes with Cauchy distributed errors, or t-student distributed errors with low degrees of freedom. The reason is that the MSE is determined mainly from the "central" values of the distribution,
and does not accommodate the tail values including the extremes, which occur during locally explosive patterns, such as bubbles and spikes. Therefore, the loss function needs to be adjusted to accommodate the extremes.
As mentioned above, in variables with infinite mean and variance, the averages are not consistent estimators, which adds to the difficulty of estimation that relies on averaging. Therefore the Mean Square Error is not an adequate loss function, and should be replaced by a log-likelihood-type of criterion.

\subsection{Impact of extreme values} 

\nin Let us examine the effect of extreme values on SGD and subsequent epochs in a simplified setup. Suppose that we have observations on the following Data Generating Process (DGP): 

\[ y_t=\theta + \epsilon_t, \qquad \epsilon_t \stackrel{\text{iid}}{\sim} \mathcal N(0,1), \qquad t=1,\ldots,T. \] 

\nin with the true parameter value denoted by $\theta^*_0$. The loss function is $\sum_{t=1}^T (y_t - \theta)^2 $. Let us define the gradient on date \(t\) : 

\[ \nabla l_t(\theta)=(-2)(Y_t-\theta), \] and denote
the initial value of the parameter by $\theta_0$ . 

\nin Assuming batch=1, i.e. online SGD training, we have the recurrence formula for the estimator: 
\begin{eqnarray*}
\hat{\theta}_t & = &  \hat{\theta}_{t-1} - \lambda_t \nabla l_t ( \hat{\theta}_{t-1}), \\
& = &  \hat{\theta}_{t-1} + 2 \lambda_t ( y_t - \hat{\theta}_{t-1}), \\
& =  & \hat{\theta}_{t-1} + \mu_t ( y_t - \hat{\theta}_{t-1}), \;\;
\mbox{with} \;\; \mu_t= 2 \lambda_t, \\
& = & (1-\mu_t) \hat{\theta}_{t-1} + \mu_t y_t \;\;
\mbox{with} \;\; \lambda_t>0.
\end{eqnarray*}

\nin  By recurrent substitutions we get: 
\begin{eqnarray*}
\hat{\theta}_1 & = & (1-\mu_1)\theta_0+\mu_1 y_1, \\
\hat{\theta}_2 & = & (1-\mu_2) \hat{\theta}_1+\mu_2 y_2 \\
& = & (1-\mu_1)(1-\mu_2)\theta_0 + (1-\mu_2)\mu_1 y_1 + \mu_2 y_2, \\\hat{\theta}_3 & = & (1-\mu_3)\hat{\theta}_2+\mu_3 y_3 = (1-\mu_1)(1-\mu_2)(1-\mu_3)\theta_0  + (1-\mu_3)(1-\mu_2)\mu_1 y_1 + (1-\mu_3)\mu_2 y_2 + \mu_3 y_3,  
\end{eqnarray*}

\nin and so on. We end up with a linear formula in \(\theta_0,y_1,\ldots,y_t\): 
\[ \hat\theta_t = A_t \theta_0 + B_{1,t} y_1 +\cdots+ B_{t,t} y_t, \] 

\nin with  $A_t=\prod_{i=1}^t (1-\mu_i)$, $B_{n,t} = [\prod_{i=n+1}^t (1-\mu_i)] \mu_n$, for $n \leq t-1$ and $B_{t,t} = \mu_t$.

Let us now examine the behavior of the estimator over subsequent epochs denoted by k=1,2,....

In epoch 1, we apply a linear filter \(F_T(\theta_0, Y_T)\) based on the observations \( Y_T= (y_1,\ldots,y_T)' \) to obtain $\hat{\theta}_{1,T}$:

\begin{equation}
\hat\theta_{1,T} = A_T \theta_0 + B_{1,T} y_1 + \cdots + B_{T,T}y_T = A_T \theta_0 + B_T' \left( \begin{array}{c} y_1 \\ \vdots \\y_T \end{array} \right) =  A_T \theta_0 + B_T'Y_T :=F_T(\theta_0, Y_T).\end{equation}

\nin In epoch 2, we apply the same filter to the same set of observations with a new initial value \(\hat{\theta}_{1,T}\), i.e. $ F_T(\hat\theta_{1,T},  Y_T)$. Then, we get:

\begin{eqnarray*}
\hat\theta_{2,T} = F_T(\hat\theta_{1,T}, Y_T) & = & 
A_T \hat\theta_{1,T} + B_{T}' Y_T \\
& = & A_T [A_T \theta_0 + B_{T}' Y_T] + B_{T}' Y_T \\
& = & A_T^2 \theta_0 + (1 + A_T) B_{T}' Y_T, 
\end{eqnarray*}

\nin and so on. After epoch \(k\), we get:

\begin{equation}
\hat\theta_{k,T} = A_T^k\theta_0 + [1 + \cdots + A_T^{k-1}] B_T' Y_T = 
A_T^k \theta_0 + (1-A_T)^{-1} (1-A_T^k) B_T' Y_T
\end{equation}

\nin Let us now examine, whether for each $k$, the estimator converges to the true parameter value $\theta_0^*$: $ \hat\theta_{k,T} \rightarrow \theta_0^*$, when $T \rightarrow \infty$, where $\theta_0^*$ is independent of epoch $k$.

\subsection{Asymptotic Analysis $T \rightarrow \infty$}

We consider $k=1$ (epoch 1) and examine the mean square convergence of 
$\hat\theta_{1,T}$, which requires $A_T \rightarrow 0$ as $T \rightarrow \infty$, and $S(T) = \sum_{n=1}^T B_{n,T}^2 \rightarrow 0$
as $T \rightarrow \infty$. These two conditions are satisfied depending on the learning rate $\mu_t$.

\nin {\bf Ex.1:} When $\mu_t =  t^{-\beta}, \beta>0$, we have the following behaviour. The term $A_T$ determines how fast the estimator forgets its initial condition $\theta_0$:
\[
A_T = \prod_{i=1}^T (1 - \alpha i^{-\beta}) = \exp\left( \sum_{i=1}^T \ln(1 - \alpha i^{-\beta}) \right)
\]
Using the Taylor expansion of $\ln(1 - \alpha i^{-\beta}) \approx -\alpha i^{-\beta}$  for large $i$, we get the variance of the estimator:

$$\text{Var}(\hat{\theta}_{1,T}) = S(T, \beta) = \sum_{n=1}^{T-1} \left[ n^{-2 \beta} \prod_{t=n+1}^T (1-t^{-\beta})^2 \right] + T^{-2 \beta}$$

\nin a) If  \(0 < \beta < 1\), then $\sum_{i=1}^T i^{-\beta} \approx \frac{T^{1-\beta}}{1-\beta}$, implying $A_T \approx \exp\left(-\frac{\alpha T^{1-\beta}}{1-\beta}\right) \longrightarrow 0$. The initial bias vanishes at a super-polynomial (exponential) rate.

Then, the variance \(\text{Var}(\hat{\theta}_{1,T}) \approx \frac{1}{2} T^{-\beta}\) tends to 0. The estimator is consistent, but its convergence is slow (under-linear), at a rate $1/T^{\beta/2}$, which is slower than the optimal square root convergence.

\nin b) When \(\beta = 1\) :  $\sum_{i=1}^T i^{-1} \approx \ln T$, implying $A_T \approx T^{-1} \longrightarrow 0$. The initial bias decays polynomially. 

\(\text{Var}(\hat{\theta}_{1,T}) = \frac{1}{T}\). This is the classical optimal case of convergence at rate \(\mathcal{O}(1/T)\)).

\nin c) If \(\beta > 1\) :  The series $\sum_{t=1}^T i^{-\beta}$ converges to a constant, implying $\lim_{T \to \infty} A_T = C_\infty > 0$, a positive bias. The effect of $\theta_0$ is never forgotten. 

\(\text{Var}(\hat{\theta}_{1,T}) \to \text{Constant} > 0\).
The step \(\mu _{t}\) decreases too fast. The algorithm stops too early and the estimator does not converge to a unique value.

\medskip
\nin The details on the derivation of these asymptotic results are given in Appendix 2. We observe that over one epoch, the learning rate plays an important role: The estimator is consistent only if $\beta \leq 1$, and the convergence rate is slow, in general. 

\medskip
If the learning rate $\mu_t = \mu$ is constant, there is no convergence over one epoch, as shown below in Example 2.

\medskip

\nin {\bf Ex. 2} Let us now consider the limiting case $\beta=0$, which yields a constant learning rate $\mu_t = \mu$, such as $\mu = 0.001$, for example. Then,
\begin{eqnarray*}
A_T & = & (1-\mu)^T \\
B_{n,T} & = & (1-\mu)^{T-n} \mu \\
B_{n,T}^2 & = & (1-\mu)^{2(T-n)} \mu^2 \\
\end{eqnarray*}
\nin where in the last term we have the sum of a geometric series.
For a constant $\mu$, the term $\sum_{n=1}^T B_{n,T}^2$ converges to a finite limit greater than 0. Hence, there is no mean-square stochastic convergence of $\hat\theta_{1,T}$ to $\theta_0^*$ in epoch $1$ with $\mu = 0.001$, for example, set by default.

\medskip
It is also interesting to see that even in the case $\beta=1$ which ensures an optimal root $T$ convergence rate examined in Example 1, the asymptotic bias of the estimator does not disappear in subsequent epochs, as shown in Example 3 below.
\medskip

\nin {\bf Ex. 3}  The asymptotic expansions are difficult to derive in general (see Appendix 2), but closed-form results are available for $\beta=1$ (Ex 1. b)
considered below.

Suppose $\beta=1$ and hence $\mu_t =1/t$. It follows that $1-\mu_t = \frac{t-1}{t}$. Then,

\begin{eqnarray*}
A_T & = & \prod_{i=1}^T \frac{i-1}{i} = \frac{2-1}{2} \frac{3-1}{3} \cdots  \frac{T-2}{T-1} \frac{T-1}{T} = \frac{1}{T} \rightarrow 0 \\
B_{n,T} & = & \mu_n (1-\mu_n) \cdots (1-\mu_T) = \frac{1}{n} \frac{n}{n+1} \cdots  \frac{T-2}{T-1} \frac{T-1}{T} = \frac{1}{T} \rightarrow 0 \\
\sum_{n=1}^T B_{n,T}^2 & = & \sum_{n=1}^T \frac{1}{T^2}= \frac{1}{T} \rightarrow 0
\end{eqnarray*}

\nin Therefore, the estimator is mean-square consistent.
However, we have $\hat\theta_{1,T} =  \frac{1}{T} \theta_0 +  \frac{1}{T} \sum_{t=1}^T Y_t = \frac{\theta_0}{T} + \bar{Y}_T$ and the estimator is asymptotically biased at order $1/T$. This bias does not disappear in subsequent epochs. To see that, consider epoch 2:

$$\hat\theta_{2,T} =  \frac{1}{T} \hat\theta_{1,T} +  \bar{Y}_T = \frac{1}{T^2} \theta_0 + (1+ \frac{1}{T})\bar{Y}_T.  $$

\nin Similarly, at epoch $k$:

\begin{eqnarray}
\hat\theta_{k,T} & = &  \frac{1}{T^k} \theta_0 + (1+ \frac{1}{T} + \frac{1}{T^2} +\cdots)\bar{Y}_T \nonumber \\
& = &  \frac{1}{T^k} \theta_0 + \Frac{1-\frac{1}{T^k}}{1-\frac{1}{T}}\bar{Y}_T 
\end{eqnarray}

\nin Overall, when $k \rightarrow \infty$ then $\hat\theta_{k,T} \approx 0 + \frac{T}{T-1}\bar{Y}_T + o(\frac{1}{T})$. We observe that the bias changed from a bias in $\frac{1}{T}$ equal to $\frac{\theta_0}{T}$ depending on the initial condition,  to a bias in $\frac{1}{T}$ equal to $E(Y_t)/T$.

\medskip
We can also describe the limiting distribution of the estimator. For $\beta=1$ (Ex 1. b)
the limiting distribution is easy to derive. Given that the convergence rate of this estimator is standard $\sqrt{T}$ and the data is normally distributed, we get the asymptotic normality of the estimator for epoch 1:

$$\sqrt{T} (\hat\theta_{1,T} - \theta_0^* - \theta_0/T) \sim N(0,1)$$

\nin and 

$$\hat\theta_{k,T} \sim N\left[\frac{1}{T^k} \theta_0, \left( \Frac{1-\frac{1}{T^k}}{1-\frac{1}{T}}\bar{Y}_T \right)^2 \frac{1}{T} \right]$$ 

\medskip
\nin for epoch $k$. 

\medskip

We may also consider the limiting distribution of the estimators in the general case of Ex 1. a, which are given in formulas (4.1) and (4.2) If $0.5<\beta <1$, we have asymptotically for epoch 1:

$$\sqrt{T^{\beta/2}} (\hat{\theta}_{1,T} - A_T \theta_0  - B_T' e \theta_0^*) \sim N(0, B_T'B_T)$$

\nin where $e$ denotes a vector of ones. For $T \rightarrow \infty$, the variance of $\hat{\theta}_{k,T}$ is obtained from the variance of $\hat{\theta}_{1,T}$ by the delta method. 
This estimator is also asymptotically normally distributed:

$$\hat{\theta}_{k,T} \sim N\left[A_T^k \theta_0 + (1-A_T)^{-1}(1-A_T^k) B_T'e\theta_0^*, \; ((1-A_T)^{-1}(1-A_T^k))^2 B_T'B_T \right]$$

\nin for epoch $k$ and has the same slow $\sqrt{T^{\beta/2}}$ rate of convergence.

\subsection{Loss Function Minimization}

We have examined so far estimator $\hat{\theta}_{k,T}$ obtained from epoch $k$ by reapplying the linear filter to the data and estimator obtained from a previous epoch. Then, 
we have generally for epoch $k$: $\hat{\theta}_{k,T} = F_T^k(\theta_0, Y_T)$, and for k=2, i.e. epoch 2: $\hat{\theta}_{2,T} = F_T(\hat{\theta}_{1,T}, Y_T)$, and so on.

Let us now consider the minimization of the MSE through the consecutive epochs $k=1,2,...$ to find an optimal choice of epoch $k^*$. We have
\begin{eqnarray*}
 \mathrm{MSE}(k)& = & \frac1T \sum_{t=1}^{T} (y_t-\hat\theta_{k,T})^2 \\
 & = & \frac1T \sum_{t=1}^{T} (y_t - \bar{y}_T)^2 + (\bar{y}_T-\hat\theta_{k,T})^2,
 \end{eqnarray*}
 
 \nin which is the sum of the variance of $y_t$ plus bias squared. Given that $\hat\theta_{k,T} = A_T^k\theta_0+ [Id + \cdots + A_T^{k-1}] B_T'\left( \begin{array}{c} y_1 \\ \vdots \\y_T \end{array} \right) = A_T^k\theta_0+ [Id + \cdots + A_T^{k-1}] B_T'Y_T $, we get:
 
\begin{eqnarray*} 
\min_k \mathrm{MSE}(k) & \iff & \min_k (\bar{y}_T-\hat\theta_{k,T})^2 \\
& \iff & \min_k (\bar{y}_T-A_T^k\theta_0 - [1 + A_T + \cdots + A_T^{k-1}] B_T' Y_T )^2\\
& \iff & \min_k \left( \left[ \bar{y}_T - \frac{B_T' Y_T}{1 - A_T} \right] - A_T^k \left[ \theta_0 - \frac{B_T' Y_T}{1 - A_T} \right] \right)^2
 \end{eqnarray*}

\nin Clearly the MSE is equal to zero when $\hat\theta_{k,T} = \bar{y}_T$. The solution of the FOC is 
$$A_T^k =  \frac{\bar{y}_T - \frac{B_T' Y_T}{1 - A_T}}{\theta_0 - \frac{B_T' Y}{1 - A_T}} = \frac{(1 - A_T)\bar{y}_T - B_T' Y_T}{(1 - A_T)\theta_0 - B_T' Y_T}$$

\nin where $\bar{y}_T $ is the maximum likelihood estimator, assuming that $\theta_0 \neq B_T' Y_T )/(1-A_T)$ 

This value needs to be positive, which is not necessarily the case depending on the data and the initial condition $\theta_0 $. Moreover, the optimal value $k^*$ needs also to be positive, which is uncertain as well, and depends on the initial condition and the data. 

We discuss these positivity conditions further in Section 4.5. For ease of exposition, we derive below the solution $k^*$ for the estimator from Ex. 3 with a constant learning rate given in formula (4.3), and point out the role of the initial condition ensuring its existence. 

\medskip
\nin {\bf Ex. 3 cont.} As mentioned earlier, the solution is 

$$k^* = \mathrm{ArgMin} \; \mathrm{MSE}(k): \hat{\theta}_{k,T} = \bar{y}_T= \bar{Y}_T$$

\nin Then, at the minimum, we have:

\begin{eqnarray*}
\frac{1}{T^k} \theta_0 + \Frac{1-\frac{1}{T^k}}{1-\frac{1}{T}} \bar{Y}_T & = & \bar{Y}_T \\   
\frac{1}{T^k} \theta_0 + \Frac{1-\frac{1}{T^k}}{1-\frac{1}{T}} \bar{Y}_T & =& 
\Frac{1-\frac{1}{T}}{1-\frac{1}{T}} \bar{Y}_T \\
\frac{1}{T^k} \theta_0 + \Frac{\bar{Y}_T}{1-\frac{1}{T}} (\frac{1}{T} - \frac{1}{T^k}) & = & 0 \\
\frac{1}{T^k} \theta_0 + \Frac{\bar{Y}_T}{1-\frac{1}{T}} \Frac{T^{k-1} -1}{T^k} & = & 0 \\
\theta_0 + \Frac{\bar{Y}_T}{1-\frac{1}{T}} (T^{k-1} -1) & = & 0 \\
T^{k-1} = 1- \Frac{\theta_0(1-\frac{1}{T}}{\bar{Y}_T}
\end{eqnarray*}
\nin so that the MSE-minimizing $k^*$ is:

$$k^* = 1+ \frac{1}{\ln T } \ln \left( 1 - \Frac{\theta_0(1-\frac{1}{T})}{\bar{Y}_T} \right)$$

\nin We observe that $k^*$ is generally greater than 1. However, this solution exists depending on the argument of the second logarithm being non-negative. This  implies a condition on the initial value: $\theta_0/ \theta_0^* \leq 1$, which can be difficult to satisfy in practice since the true parameter value is unknown. This suggests that experimenting with multiple choices of initial values in practice may be helpful.

\subsection{Process with an extreme value}

\medskip

\nin Suppose now that the process contains an extreme value that occurs at time $t=10$:

\[ y_t=\theta+ \delta \mathbf{1}_{t=10}+ \epsilon_t, \qquad \epsilon_t \stackrel{\text{iid}}{\sim} \mathcal N(0,1), \qquad t=1,\ldots,T, \qquad\epsilon_{10}^* = \epsilon_{10} + \delta\] 

\nin with the true parameter value $\theta^*_0$ as before. The loss function is now:

\[
L_t= \sum_{t=1}^T (y_t - \theta^* - \mathbf{1}_{t=10} \delta^*)^2
\]

\nin For all $t < 10$, the recurrent estimator formula is:

\begin{eqnarray*}
\hat{\theta}_t & = &  \hat{\theta}_{t-1} -\lambda_t \nabla l_t ( \hat{\theta}_{t-1}), \\
& = & (1-\mu_t) \hat{\theta}_{t-1} + \mu_t y_t, \;\; t=1,...,9
\end{eqnarray*}

\nin At $t=10$, we have a change in the gradient

$$ \nabla l_{10}(\theta)=(-2)(Y_{10}(\delta) - \theta) =  (-2)(Y_{10} + \delta^* - \theta)$$ 

\nin with $y_{10} = \theta^* + \epsilon_{10}^*$.

We consider the linear filter considered in Section 4.3: $F_T(\theta_0, Y_T) = A_T \theta_0 + \sum_{t=1}^T B_{t,T} y_t$.
Recall that without the shock $\delta$, the inputs to filter $F_T(\theta_0, Y_T)$ are $y_1,...,y_T$ and the initial value $\theta_0$. Then, the output is $\hat{\theta}_{1,T}$: $\hat{\theta}_{1,T} = F_T( y_1,...,y_T, \theta_0)$ in epoch 1.
In epoch 2, without the shock, the inputs are $\hat{\theta}_{1,T}$ as the new starting value and the observations. The output is $\hat{\theta}_{2,T}$, and so on.

With a shock $\delta^*$ at time $t=10$, the inputs to the filter are the initial condition $\theta_0$ and $y_1(\delta^*)....y_T(\delta^*)$. As the output from epoch 1, we get  $\hat{\theta}_{1,T}(\delta^*)$ . Since the filter is linear, we have in epoch 1: $B_{10,T} \delta^* = \hat{\theta}_{1,T}(\delta^*) - \hat{\theta}_{1,T}$, so that due to the shock:
$$\hat{\theta}_{1,T}(\delta^*) = \hat{\theta}_{1,T} + B_{10,T} \delta^* $$

\nin In epoch 2, this quantity is used as the starting value. Then, filter $F_T(.)$ produces the following output:
$$\hat{\theta}_{2,T}(\delta^*) = \hat{\theta}_{2,T} + A_T B_{10,T} \delta^*  + B_{10,T} \delta^* $$

\nin Using the same symbols as before, we can examine the behavior of the estimator through epochs $k=1,2,....$. In epoch 1, we have:

$$\hat\theta_{1,T} = A_T \theta_0 + B_{1,T} y_1 + \cdots + B_{T,T}y_T + B_{10,T} \delta = A_T \theta_0 + B_T' Y_T +  B_{10,T} \delta . $$

\nin In epoch 2, we apply the filter $ F(\hat\theta_{1,T}, Y_T)$ with a new initial value \(\hat{\theta}_{1,T}\). Then, we get 
$$\hat{\theta}_{2,T} = A_T^2 \theta_0 + (1 + A_T) B_T' Y + A_T B_{10,T} \delta = A_T^2 \theta_0 + (1 + A_T) B_T' Y_T + (1+A_T) B_{10,T} \delta $$

\nin and so on. In epoch \(k\), we obtain:
\begin{eqnarray*}
 \hat\theta_{k,T} & = & A_T^k\theta_0 + (1+A_T+ \cdots +A_T^{k-1}) B_T' Y_T + (1+A_T+ \cdots +A_T^{k-1}) B_{10,T} \delta \\
&= & A_T^k \theta_0 + \frac{1 - A_T^k}{1 - A_T} B_T' Y_T + \frac{1 - A_T^k}{1 - A_T} B_{10,T} \delta 
\end{eqnarray*}

\nin The middle term converges to $\theta_0^*$, the first one to zero, and the last one is a bias term. The timing of the shock has an effect on its size. The more epochs, i.e. the larger $k$, the less effect the last term has on $\hat\theta_{k,T}$.

The term \(B_{10,T} \delta\) modifies the previous asymptotic results on the convergence of the estimator. This term represents a systematic bias (due to a constant derivative \(\delta \)). Because the fraction \(\frac{1-A_{T}^{k}}{1-A_{T}}\) depends on the behaviour of \(A_{T}\), the asymptotic impact of \(\delta \) will amplify itself or dissipate depending on the value  \(\beta \) in the learning rate which determines \(A_{T}\) and \(B_{10,T}\).

\medskip
\nin {\it Minimization of MSE Loss Function}

Let us now examine the MSE loss function, written as a function of epoch $k$  
$$
 \mathrm{MSE}(k) =  \frac1T \sum_{t=1}^{T} (y_t-\hat\theta_{k,T} - \mathbf{1}_{t=10} \delta)^2 
$$

\nin We aim to minimize the MSE loss function, with respect to $k$ to find the optimal choice of epoch $k^*$ assuming a shock at time t=10:
\[
\mathrm{MSE}(k) = \left( \bar{y}_T - \hat{\theta}_{k,T} - \mathbf{1}_{t=10} \delta \right)^2
\]
The global minimum is achieved when the expression inside the square equals zero:
\begin{equation}
\hat{\theta}_{k,T} = \bar{y}_T - \mathbf{1}_{t=10} \delta
\label{eq:opt_cond}
\end{equation}

\nin To find $k^*$ that minimizes the MSE(k), we proceed as follows. Recall the expression for $\hat{\theta}_{k,T}$ derived above:
\begin{eqnarray*}
\hat{\theta}_{k,T} & = & A_T^k \theta_0 + \frac{1 - A_T^k}{1 - A_T} B_T' Y + \frac{1 - A_T^k}{1 - A_T} B_{10,T} \delta \\    
& = & A_T^k \left( \theta_0 - \frac{B_T' Y + B_{10,T} \delta}{1 - A_T} \right) + \frac{B_T' Y + B_{10,T} \delta}{1 - A_T}
\end{eqnarray*}

\nin Substituting this expression into the optimality condition \eqref{eq:opt_cond} gives:
\[
A_T^k \left( \theta_0 - \frac{B_T' Y + B_{10,T} \delta}{1 - A_T} \right) + \frac{B_T' Y + B_{10,T} \delta}{1 - A_T} = \bar{y}_T - \mathbf{1}_{t=10} \delta
\]
and hence
\[
A_T^k \left( \theta_0 - \frac{B_T^\top Y + B_{10,T} \delta}{1 - A_T} \right) = \bar{y}_T - \mathbf{1}_{t=10} \delta - \frac{B_T^\top Y + B_{10,T} \delta}{1 - A_T}.
\]
The solution for $A_T^k$ is:
\[
A_T^k = \frac{\bar{y}_T - \mathbf{1}_{t=10} \delta - \frac{B_T' Y + B_{10,T} \delta}{1 - A_T}}{\theta_0 - \frac{B_T' Y + B_{10,T} \delta}{1 - A_T}}
\]

\nin To simplify the fraction, we multiply both the numerator and the denominator by $(1 - A_T)$:
\[
A_T^k = \frac{(1 - A_T)(\bar{y}_T - \mathbf{1}_{t=10} \delta) - (B_T' Y + B_{10,T} \delta)}{(1 - A_T)\theta_0 - (B_T' Y + B_{10,T} \delta)}
\]

\nin Taking the natural logarithm on both sides yields:
\[
k \ln(A_T) = \ln \left( \frac{(1 - A_T)(\bar{y}_T - \mathbf{1}_{t=10} \delta) - (B_T^\top Y + B_{10,T} \delta)}{(1 - A_T)\theta_0 - (B_T^\top Y + B_{10,T} \delta)} \right)
\]
and the optimal value $k^*$ is:
\[
k^* = \frac{1}{\ln(A_T)} \ln \left( \frac{(1 - A_T)(\bar{y}_T - \mathbf{1}_{t=10} \delta) - (B_T' Y + B_{10,T} \delta)}{(1 - A_T)\theta_0 - (B_T' Y + B_{10,T} \delta)} \right)
\]

\nin The above formula illustrates the effect of the shock, and the importance of its size and sign.
For a valid, real-valued solution $k^*$ to exist, the following conditions must hold:

\begin{itemize}
    \item[i)] The argument inside the second logarithm must be strictly positive, while it depends on the shock size and sign.
    \item[ii)] $A_T \neq 1$ (which is true since $A_T = \prod (1-\mu_i) < 1$ for $\mu_i > 0$).
    \item[iii)] If $k$ is constrained to be an integer representing the number of epochs, the optimal discrete $k$ has to be rounded up or down, depending on which yields a lower value for $MSE(k)$.
For this condition to hold, we need the numerator and denominator of the fraction to be of the same sign, which is uncertain, depending on the sign of the shock, with either: $\bar{y}_T - \mathbf{1}_{t=10} \delta - \frac{B_T' Y + B_{10,T} \delta}{1 - A_T} >0$ and $\theta_0 - \frac{B_T' Y + B_{10,T} \delta}{1 - A_T}>0$ and their ratio being strictly positive, or both these terms being negative, with their ratio being strictly positive.
Moreover, the  shock term does not disappear with an increasing number of epochs. 
The timing of the shock matters for the bias: the closer to $T$, the larger its effect.
\end{itemize}
\bigskip

\nin {\it Asymptotic Analysis of $k^*$ as $T \to \infty$}

Let us now consider an infinitely large sample $T \to \infty$. This scenario sheds light on the practice of "pooling" in applied research illustrated later in Section 5.3. We expect that for an infinitely large sample, the optimal number of epochs $k^*$ should tend to zero.

To understand how the optimal parameter $k^*$ changes when the sample size $T$ becomes infinitely large, we rewrite it as:
\begin{equation}
k^* = \frac{\ln(N_T)}{\ln(A_T)}
\end{equation}
where $N_T$ is the ratio:
\[
N_T = \frac{(1 - A_T)(\bar{y}_T - \mathbf{1}_{t=10} \delta) - (B_T^\top Y + B_{10,T} \delta)}{(1 - A_T)\theta_0 - (B_T^\top Y + B_{10,T} \delta)}
\]
\nin Its asymptotic behaviour depends on the learning rate coefficient $\beta$ in $\mu_t = \alpha t^{-\beta}$ and on the shock $\delta$.

\nin When $T \to \infty$ then by the Law of Large Numbers, $\bar{y}_T \to \theta_0^*$, the sequence $B_T' Y_T$ is a weighted moving average tracking $\theta^*_0$. As $T \to \infty$, if the system is stable, $(B_T' Y + B_{10,T} \delta) \to (1 - A_T)\theta^*_0$. Substituting these limits simplifies the fraction to:
    \[
    N_T \approx \frac{(1-A_T)(\theta^*_0 - \mathbf{1}_{t=10} \delta) - (1-A_T)\theta^*_0}{(1-A_T)\theta_0 - (1-A_T)\theta^*_0} = \frac{-\mathbf{1}_{t=10} \delta}{\theta_0 - \theta^*_0}
    \]
\nin Thus, as $T \to \infty$, $N_T$ in the numerator of (4.5) approaches a stable, finite, non-zero constant $C_N$  determined by the shock: $C_N=\frac{-\mathbf{1}_{t=10} \delta}{\theta_0 - \theta^*_0}$, implying $\ln(N_T) = O(1)$, as long as the argument of the logarithm is positive.

Then, the effect of the learning rate on the asymptotic behavior of $k^*$ is determined by the denominator $\ln(A_T)$. We consider the learning rate $\mu_t$ with the following coefficients:

\nin a) $0 < \beta < 1$
Under this learning rate, the memory term drops super-polynomially:
\[
A_T \approx \exp\left(-\frac{\alpha T^{1-\beta}}{1-\beta}\right) \implies \ln(A_T) \approx -\frac{\alpha}{1-\beta} T^{1-\beta}
\]
Since $\ln(N_T)$ remains a constant $O(1)$, we substitute it into formula (4.5) of $k^*$:
\[
k^* = \frac{O(1)}{-\frac{\alpha}{1-\beta} T^{1-\beta}} = O\left(T^{-(1-\beta)}\right)
\]
As $T \to \infty$, $k^* \to 0$. The learning rate increases fast enough to overwrite the initial condition and shock effect so efficiently that no multiple epochs ($k>1$) are required to minimize the error when the sample size tends to infinity.

\medskip
\nin b) $\beta = 1$ (Optimal Robbins-Monro step size)
In this setting, the memory term decays polynomially:
\[
A_T \approx T^{-\alpha} \implies \ln(A_T) \approx -\alpha \ln(T)
\]
Substituting this value yields the optimal count of epochs:
\[
k^* = \frac{O(1)}{-\alpha \ln(T)} = O\left(\frac{1}{\ln(T)}\right)
\]
\nin As $T \to \infty$, $k^* \to 0$ logarithmically slow. Hence, multiple epochs are needed even for an infinitely large sample size. The optimal number of epochs decreases slowly to 0 while repeated processing of the infinitely large data set leads to convergence.

\medskip
\nin c) $\beta > 1$ When the learning rate shrinks too fast, $A_T$ converges to a  positive constant $C_\infty$:
\[
\lim_{T\to\infty} A_T = C_\infty \in (0, 1) \implies \ln(A_T) \to \ln(C_\infty) = O(1)
\]
which implies:
\[
k^* \to \frac{\ln(C_N)}{\ln(C_\infty)} = O(1)
\]
\nin The optimal number of epochs $k^*$ is a constant, finite and greater than zero. Because the individual update rate size $\mu_t$ is too fast to eliminate bias, a fixed positive number $k^* > 0$ of epochs is needed even when the data set is infinitely large.

\bigskip
The analysis presented so far concerned a Gaussian process. When a process with a heavy-tailed distribution is considered, the non-convergence of sample averages to the theoretical moments, mentioned earlier, complicates the analysis further. Then, the choice of initial conditions becomes more impactful, and finding optimal epoch numbers becomes more difficult. Moreover, when sample averages are not consistent estimators of the theoretical moments, the MSE is no longer an adequate training loss function, and likelihood-type loss functions, which can be consistently estimated, have to be used instead. This issue is illustrated and examined further through simulations in Section 5.4.

\setcounter{equation}{0}\def\theequation{5.\arabic{equation}}
\section{Simulation Study}

This Section examines how well the different types of neural nets can predict extreme values that occur during a bubble. 

All training processes in this paper were conducted on a desktop workstation equipped with a 13th Gen Intel Core i7-13700KF CPU, 31.83 GB of system RAM, and an AMD Radeon RX 7900 XTX GPU. The AMD GPU was accessed through PyTorch's CUDA device interface, using \textit{Python 3.12} with \textit{ROCm-enabled PyTorch 2.9.1+rocmsdk20260116}. To ensure stable execution for recurrent neural network models on this device, the MIOpen/cuDNN backend was disabled during training, while tensor computations were still performed on the GPU. The comparison of solo-trained and pooled-training times for each algorithm is provided in Table \ref{tab:pooled time} of the Appendix.

\subsection{White noise process with extreme values}

Let us examine the predictions of a Cauchy distributed white noise process, which is an example of a noise process without finite marginal moments. In this setup, the averages are inconsistent estimators of the theoretical moments, and we expect convergence problems of the weight and bias parameters. We simulate a series of white noise with 2000 observations from a standard Cauchy distribution: $y_t \overset{\mathrm{i.i.d}}{\sim}\operatorname{Cauchy}(0,1)$. The first 90\% of the observations are used as the training set, and the remaining 10\% as the test set for out-of-sample evaluation. 

The objective is one-step-ahead prediction. We conduct experiments separately on input sequence length (time steps = 1, 5, 10), number of hidden units (unit = 1,2,4), batch size (batch = 1, 16,32,64, 128), and number of epochs (epochs = 1, 10, 100, 1000, 5000,10000). During each experiment, all settings except the target hyperparameter are held constant throughout the training process\footnote{The learning rate for all experiments is set to $0.001$.} .

To compare the model performance using a robust measure instead of the non-convergent RMSE, we consider the Root Median Square Error (RMedSE)\footnote{defined as: $RMedSE = \sqrt{ med ((\hat{y}_t - y_t)^2)}$
where $\hat{y}_t$ is the forecast conditional on $\hat{\theta}$ and $Y_{t-1}$ in test set.} computed over the test set. The results are illustrated in Tables \ref{t3} to \ref{t4}.

\begin{table}[H]
\centering
\caption{Forecast Performance for the Cauchy White Noise Batch-size Experiment}
\label{tab:cauchy_batch_size}
\begin{tabular}{ccc}
\toprule
 Batch size  & RMedSE \\
\midrule
 1    & \textbf{2.965} \\
 16   & 5.884 \\
 32   & 5.974 \\
 64   & 6.387 \\
 128  & 6.338 \\
\bottomrule
\end{tabular}
\label{t3}
\caption*{\small \textit {The number of hidden units and time step are both fixed at 1. The number of epochs are fixed at 10,000}}
\end{table}

\begin{table}[H]
\centering
\caption{Forecast performance for the Cauchy White Noise Epoch Experiment. }
\label{tab:cauchy_epochs}
\begin{tabular}{lcc}
\toprule
Epochs  & RMedSE \\
\midrule
1       & \textbf{2.471} \\
10      & 2.507 \\
100     & 2.967 \\
1,000   & 2.965 \\
5,000   & 2.965 \\
10,000  & 2.965 \\
\bottomrule
\end{tabular}
\label{t4}
\caption*{\small \textit {The number of hidden units, time step, and batch size are all fixed at 1.}}
\end{table}

These results indicate that out-of-sample prediction performance may not improve with an increased batch size or number of epochs in the training process.

Figure 6 below illustrates a trajectory of the time series of Cauchy white noise (blue line) and the in-sample and out-of-sample one-step ahead predictions (red line) in panels a and b, respectively.

\begin{figure}[H]
    \centering
        \caption{LSTM Prediction of Cauchy White Noise Trajectory}
    \begin{subfigure}[b]{0.48\textwidth}
        \centering
        \includegraphics[width=\linewidth]{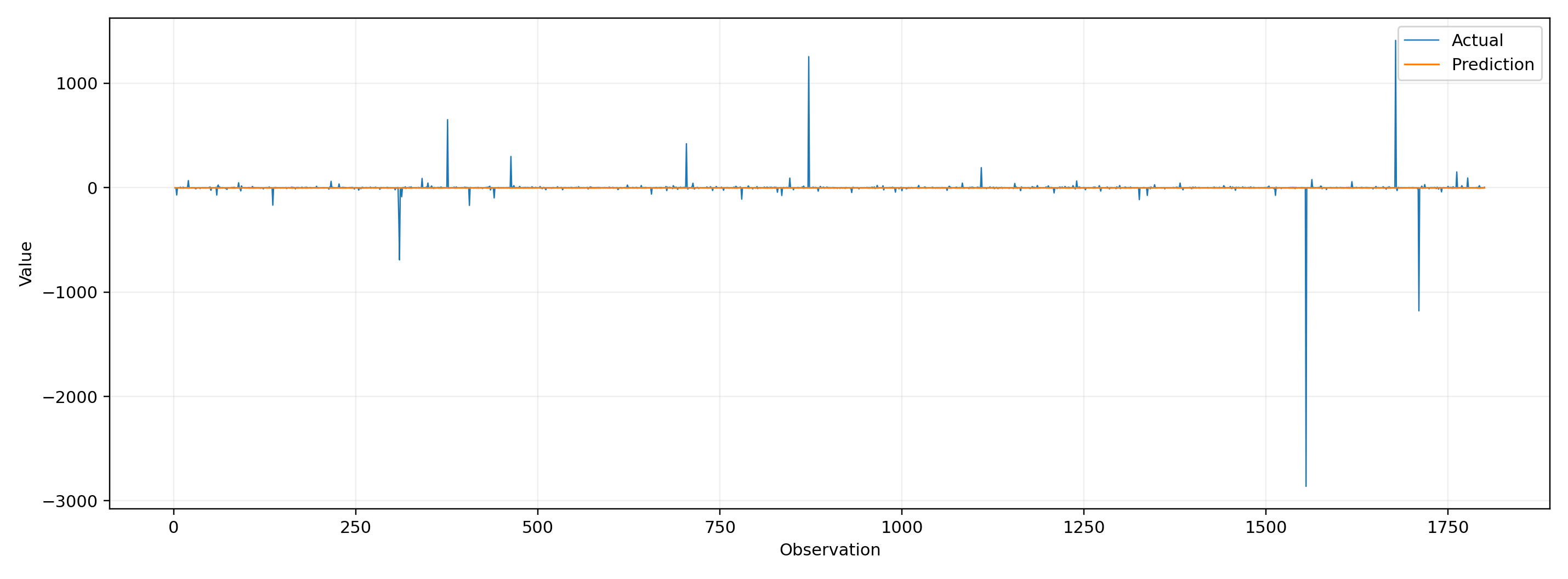}
        \caption{In-sample Fitted Values}
        \label{wnin}
    \end{subfigure}
    \hfill
    \begin{subfigure}[b]{0.48\textwidth}
        \centering
        \includegraphics[width=\linewidth]{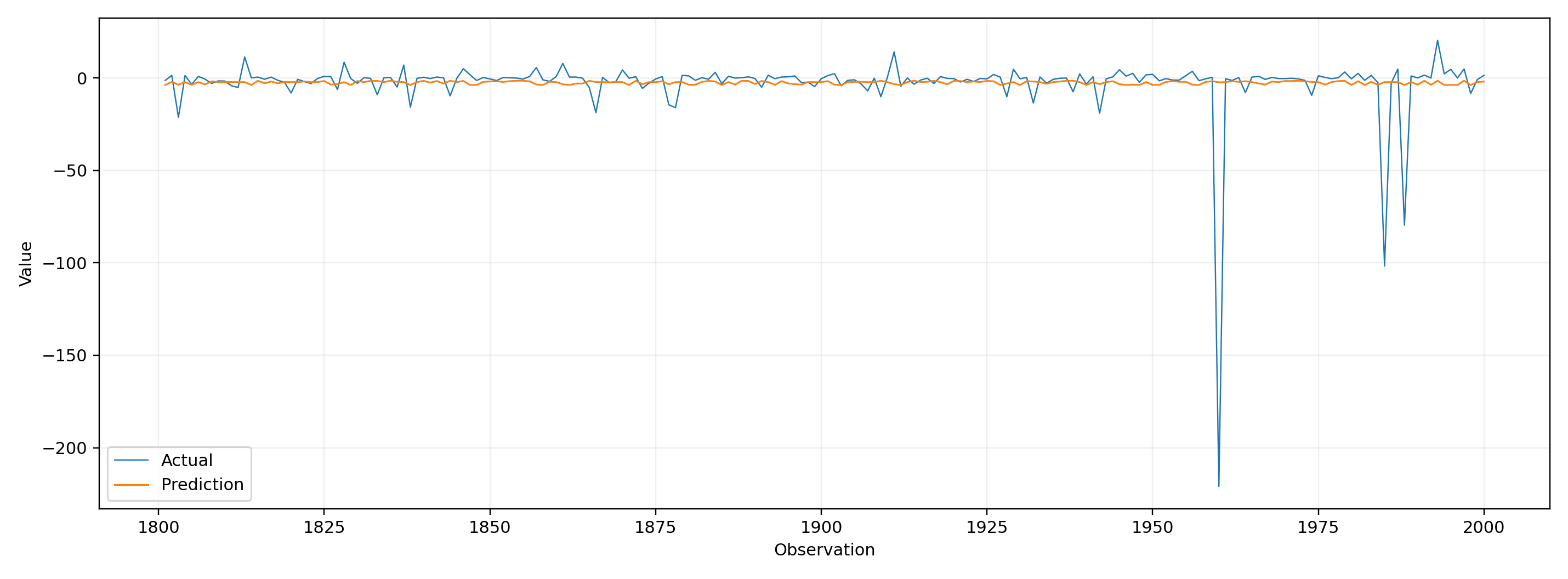}
        \caption{Out-of-sample Prediction Values}
        \label{wnfig}
    \end{subfigure}

    \label{fig:combined}
\end{figure}

We observe that the predictions of extreme values are rather poor both in- and out-of-sample.

\subsection{MAR(1,1) processes with locally explosive patterns}

We consider MAR(1,1) processes with the coefficients $\phi=0.3$, $\psi=0.8$
and t-Student distributed errors with degrees of freedom from 1 to 5. The t-Student with 1 degree of freedom is the Cauchy error distribution with no finite marginal moments.

We carry out two experiments with these processes divided into the following two groups:

\medskip 

\nin Group 1: bubble in the training set and in the test set

\medskip
\nin Group 2: no bubble in the training set but in the test set.

Each simulated path is of length 400 and each experiment is replicated 150 times. Group 1 contains 100 trajectories, and Group 2 contains 50 trajectories for each degree of freedom from 1 to 5. For each trajectory, the first 200 observations (\(t = 1,\ldots,200\)) form the training set, and the remaining 200 observations (\(t = 201,\ldots,400\)) form the test set. In this section, we estimate each trajectory separately.

Depending on the degree of freedom of the error distribution, the bubbles are large and infrequent in the Cauchy distributed processes and smaller and relatively frequent in the t-Student distributed processes with larger numbers of degrees of freedom. The bubbles are the most frequent and smallest in the process with t(5) distributed errors.

To ensure that there are no bubbles in the training set, we select only processes with the maximum realization below the 98th theoretical percentile, approximated from a large number of simulations. To ensure that there is a bubble in the testing set, we select the trajectories with at least one value above the same threshold \footnote{ The threshold is set equal to the mean of the empirical 98\% quantiles of each distribution with the degrees of freedom from 1 to 5 replicated 10000 times.}.

\nin We use the tuning parameters given in Table \ref{hyper}:

\begin{table}[htbp]
\centering
\caption{Hyperparameters for Solo-trained Forecasting Models}
\begin{tabular}{lrrrrl}
\hline
Model & Units & Batch & Reg. & Epochs & Parameters \\
\hline
FNN           & 128 & 32 & 0.001 & 2000 & 17,409  \\
RNN           & 128 & 128 & 0.001 & 1000 & 16,897  \\
LSTM          & 64 & 32 & 0.0001 & 2000 & 17,217  \\
GRU           & 128 & 32 & 0.0001 & 1000 & 50,433  \\
BiLSTM        & 128  & 64 & 0.0001 & 2000 & 134,401  \\
\hline
\end{tabular}
\label{hyper}
\end{table}

We set the lookback time to $5$. The last $10\%$ of observations in the test sets are used as a validation set.  The tuning hyperparameters, reported in Table \ref{hyper}, for each model are chosen by running a grid search on a  subset of simulated trajectories.

\begin{figure}[H]
    \centering
    \caption{Group 1 Test Set Forecast Example}

    \begin{subfigure}[b]{\textwidth}
        \centering
        \includegraphics[height=4cm]{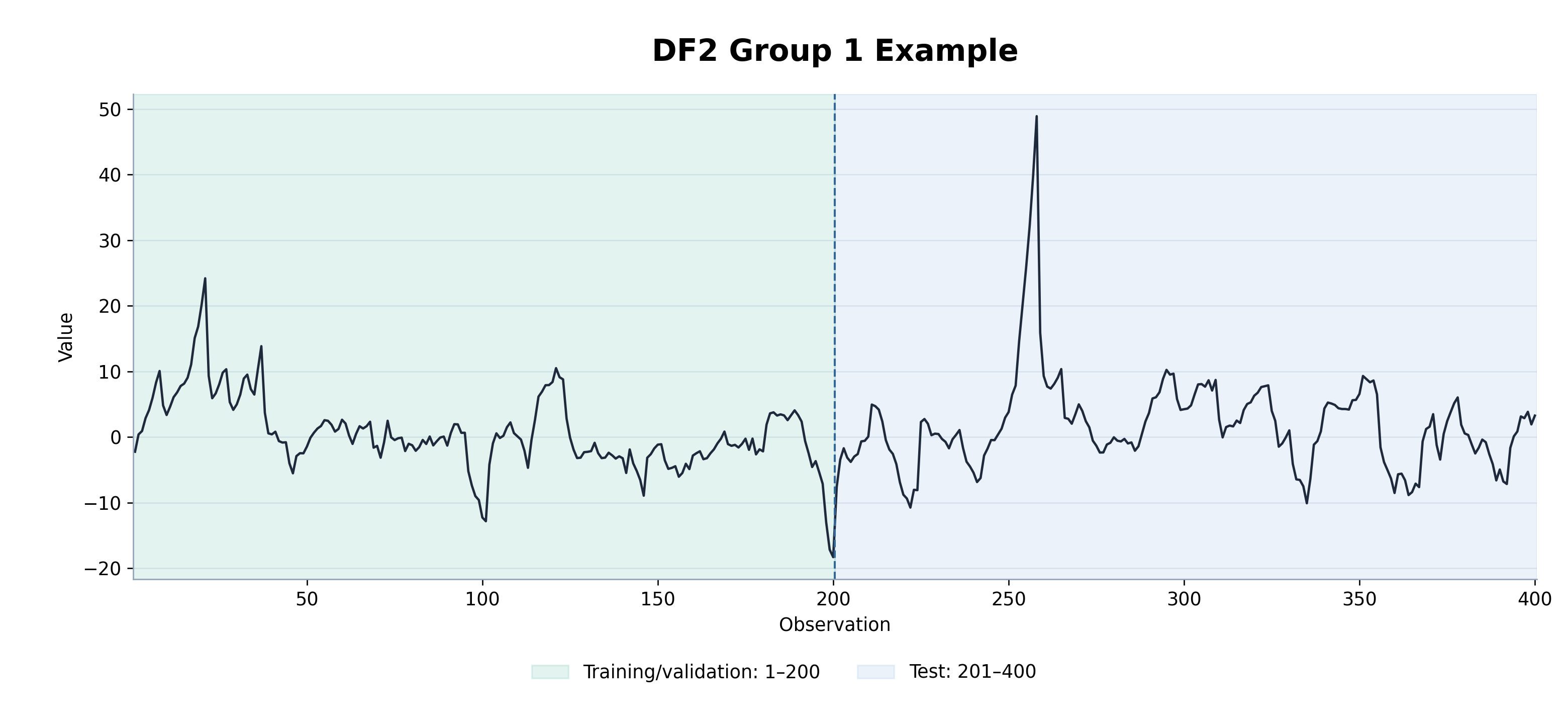}
        \caption{DF2 Group 1 series Example}
        \label{fig:df2_series}
    \end{subfigure}

    \vspace{0.5cm}

    \begin{subfigure}[b]{\textwidth}
        \centering
        \includegraphics[height=10cm]{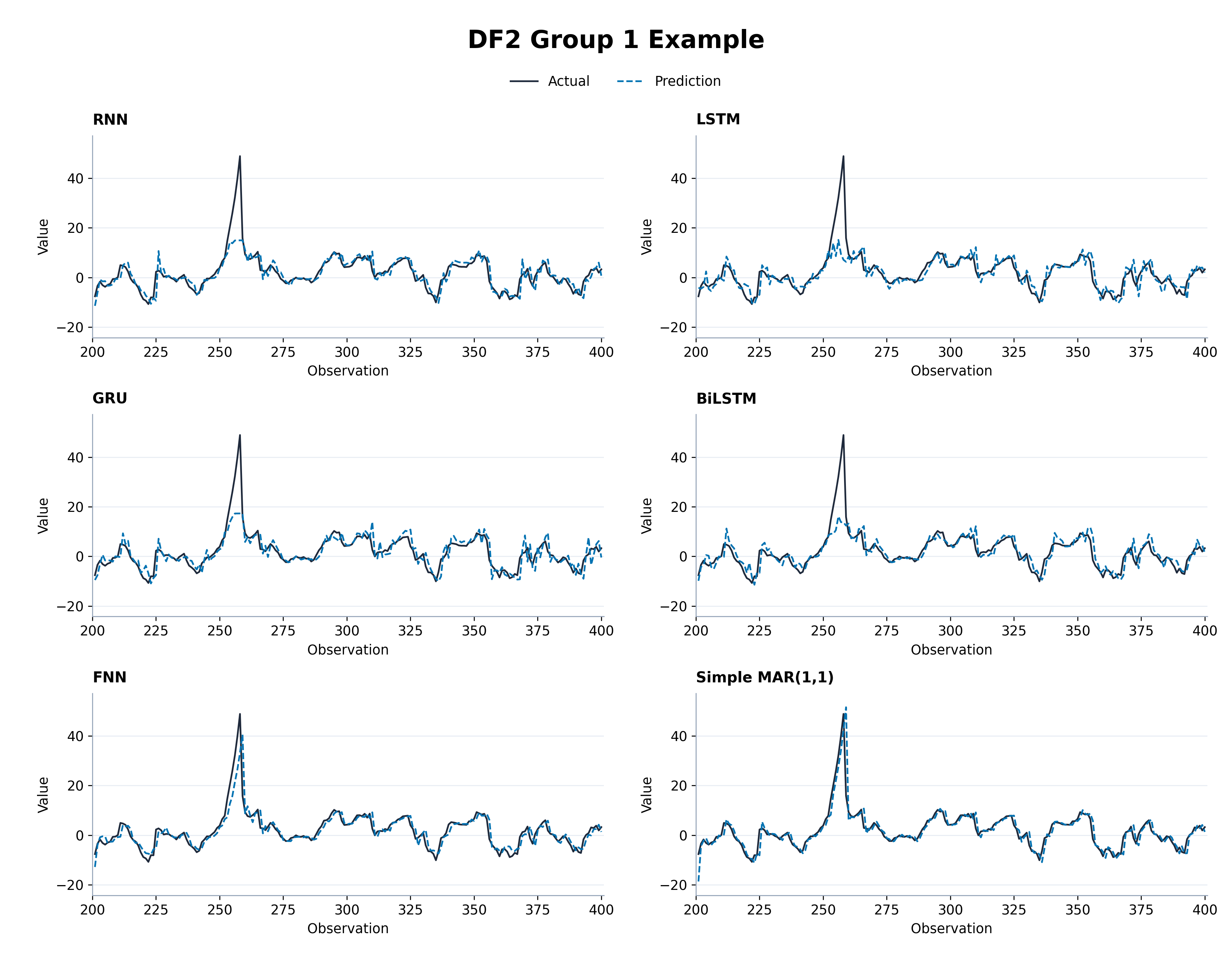}
        \caption{Test Set Forecast Results}
        \label{df2_fore}
    \end{subfigure}

    \label{df2group1}
\end{figure}

\begin{figure}[H]
    \centering
    \caption{Group 2 Test Set Forecast Example}

    \begin{subfigure}[b]{\textwidth}
        \centering
        \includegraphics[height=6cm]{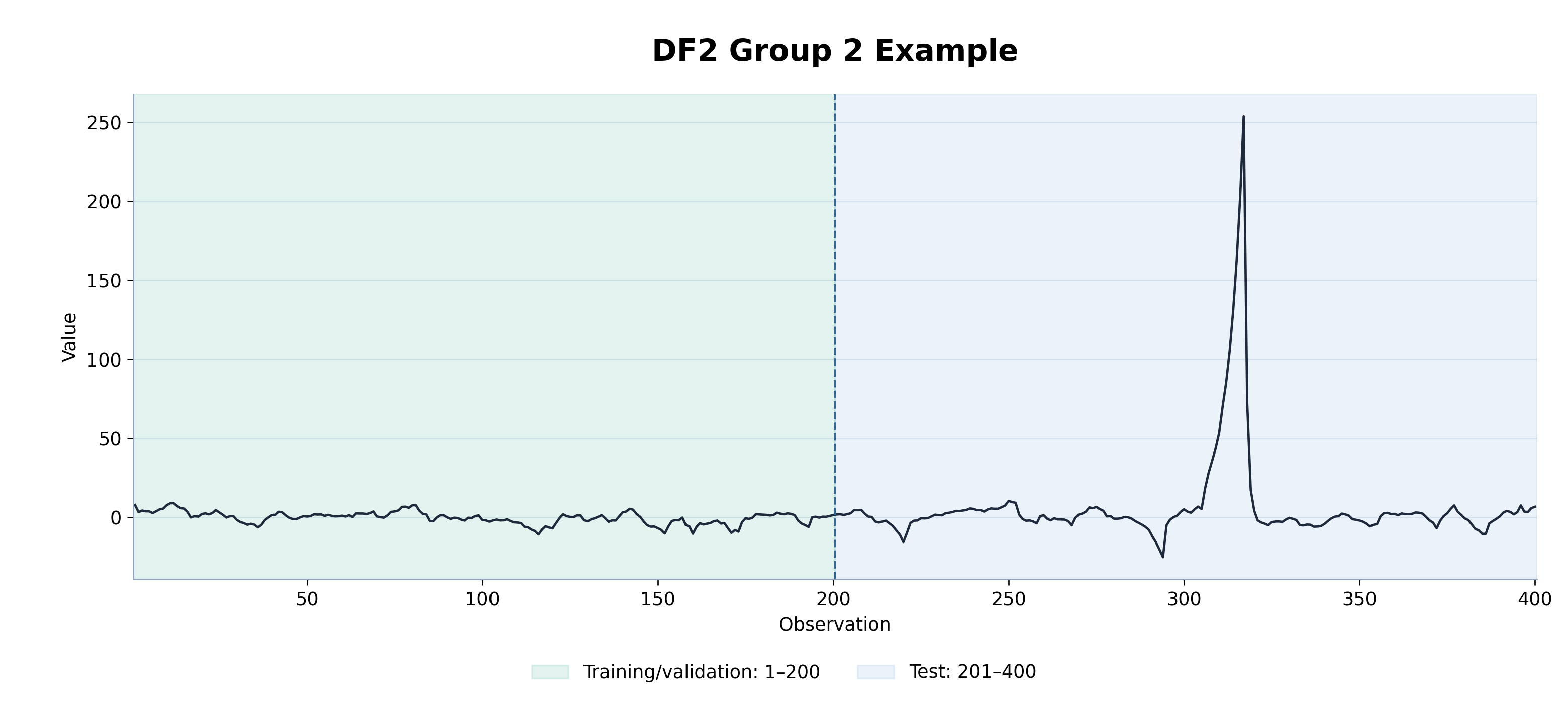}
        \caption{DF2 Group 2 series Example}
        \label{fig:df2_series2}
    \end{subfigure}

    \vspace{0.5cm}

    \begin{subfigure}[b]{\textwidth}
        \centering
        \includegraphics[height=12cm]{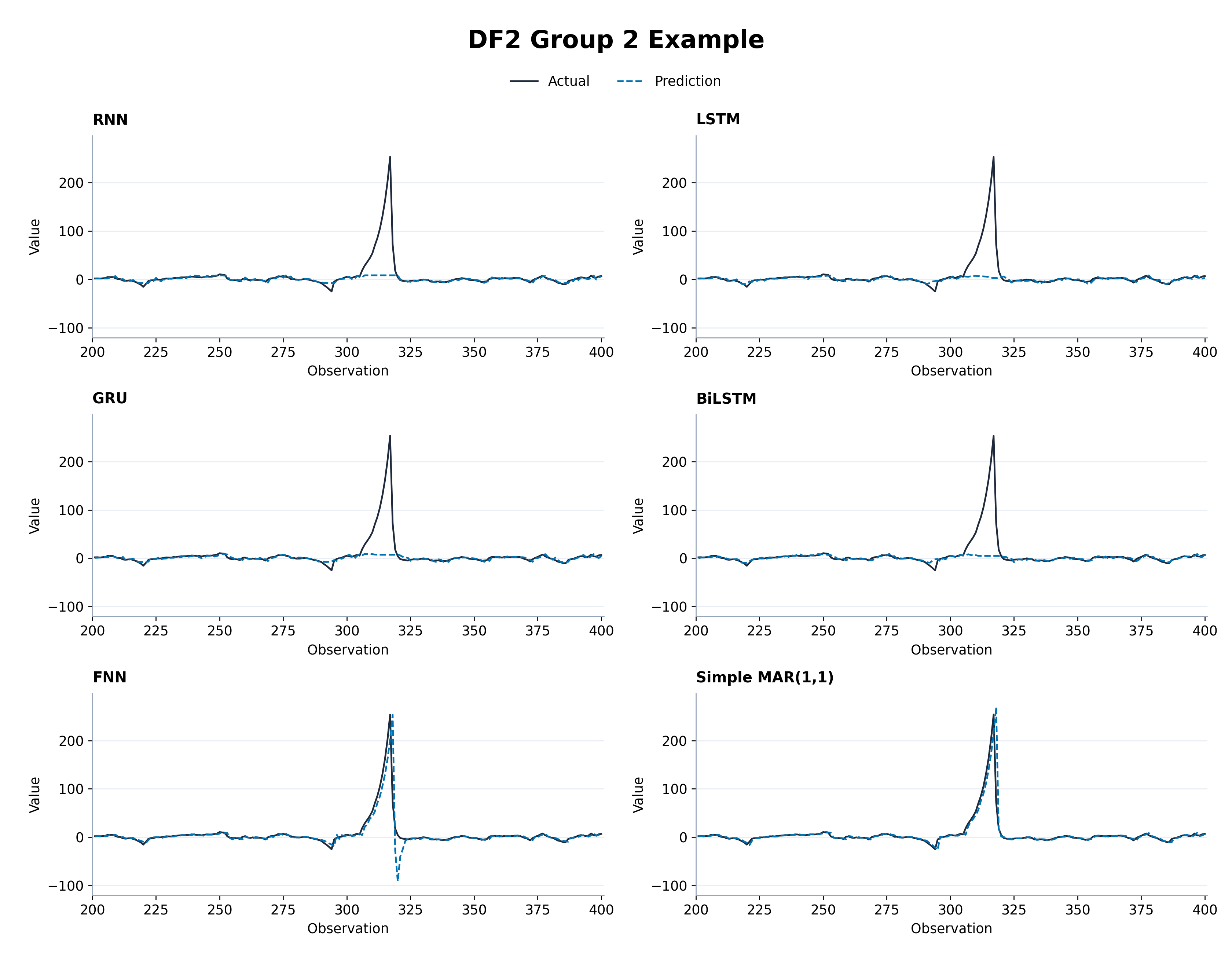}
        \caption{Test Set Forecast Results}
        \label{df2_fore2}
    \end{subfigure}
    \label{df2group2}
\end{figure}

Figures \ref{df2group1} and \ref{df2group2} demonstrate typical forecast results obtained from Group 1 and Group 2 processes using the MAR(1,1) model with t-student distributed errors with degrees of freedom 2. Each last panel reports the benchmark forecast obtained from formula (2.4) labeled as "Simple MAR(1,1)".

We observe that FNN is the best performing method both in groups 1 and 2. Intuitively, this method involves the least amount of data rescaling by activation functions over limited sets, such as [0,1], or [-1,1]. 
While the RNN and LSTM can both be written as nonlinear AR(infinity) process, the FNN is Markov (or finite-order Markov if several neighboring observations are stacked together as input). Since MAR(1,1) is second-order Markov, this also explains why the FNN performs better than RNN/LSTM. 

Moreover, we find that the presence of a bubble in the training set does not ensure the ability of the model to predict extreme values.

\medskip

\begin{figure}[H]
    \centering
    \caption{Test Sets Forecast RMSE for Bubble Episodes Distribution}
    \includegraphics[width=0.8\linewidth]{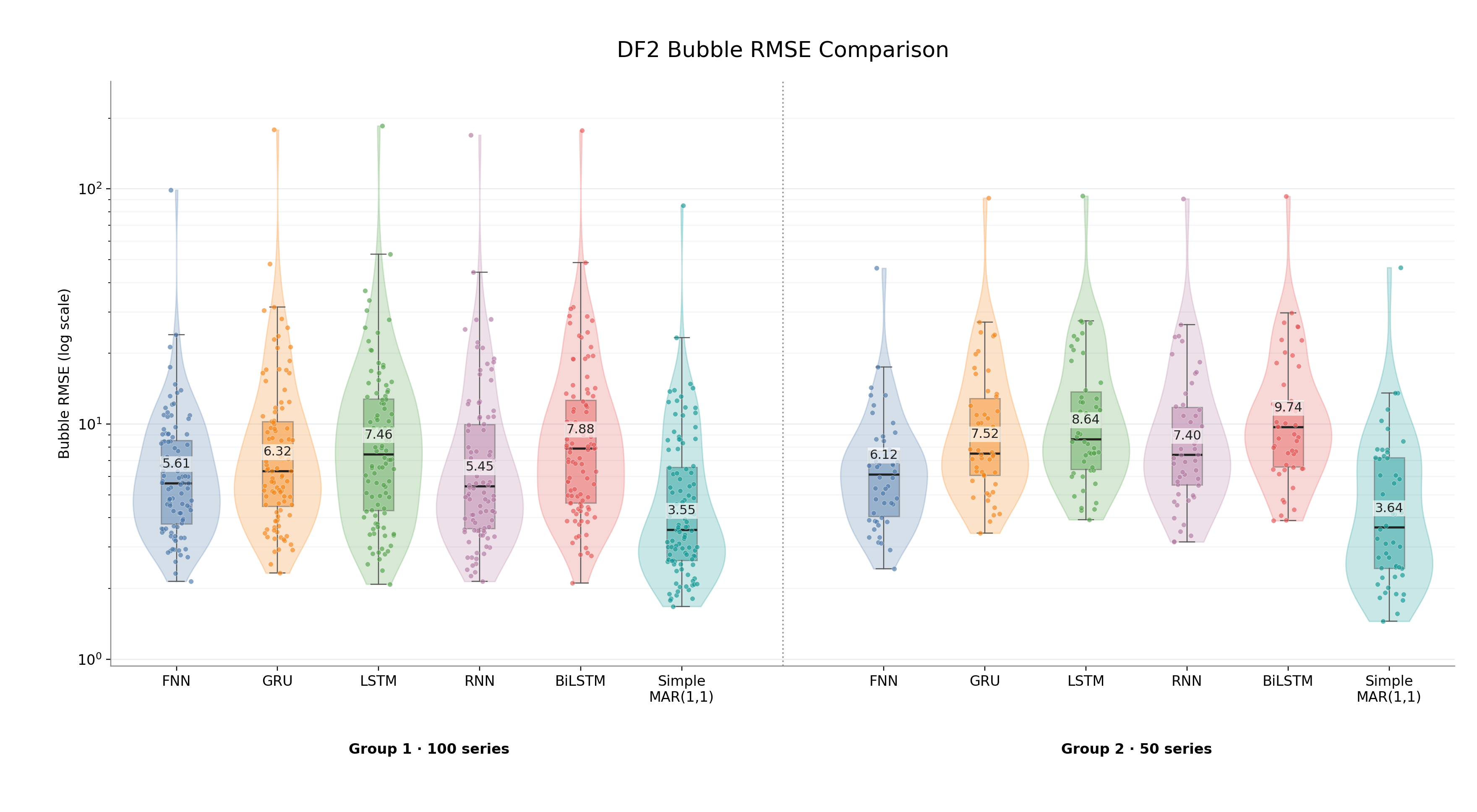}
        			\vspace{0.1cm}
			\hrule
			\vspace{0.1cm}
    \caption*{The horizontal bars and values represent the median bubble RMSE for each group and algorithm.} 
    \label{rmse dis}
\end{figure}

Figure \ref{rmse dis} illustrates the test set-based RMSE\footnote{defined as $RMSE = \sqrt{\frac{1}{n}\sum_{t=1}^n((\hat{y}_t|Y_{t-1}, \hat{\theta}) - y_t)^2|Y_{t-1}) \id_{y_t \geq q_{0.9} }}$
where $n$ is the number of forecast values in the test set, $Y_{t-1}$ denotes the past values and $\hat{y}_t$ is the forecast in test set} of bubble forecast from the recurrent nets and compares the target algorithms with the RMSE of sample MAR(1,1) (formula (2.4)) as a benchmark. The bubble RMSEs are calculated only from the top 10\% absolute values of observations in the test sets and their corresponding forecast values.

We find that the forecast performance of all nets is worse than the simple MAR(1,1)-forecast based on formula (2.4). The RNN performs relatively better in Group 1, and the FNN has the lowest median RMSE in Group 2.

\subsection{Pooled Mechanism}

Training each trajectory independently using training sets of 200 observation is less effective due to a large number of parameters in most RNN models. We concatenate all training sets together (per group, per degree of freedom) to train one pooled model. For each degree of freedom from 1 to 5, the training set has $200\times100 = 20,000$ observations for Group 1, and $200\times50 = 10,000$ observations for Group 2. The hyperparameters used in each model are reported in Table \ref{tab:model_tuning_parameters_simple} of the Appendix.
The results are illustrated in Figures 10-11.

\begin{figure}[H]
    \centering
    \caption{}
    \includegraphics[width=0.9\linewidth]{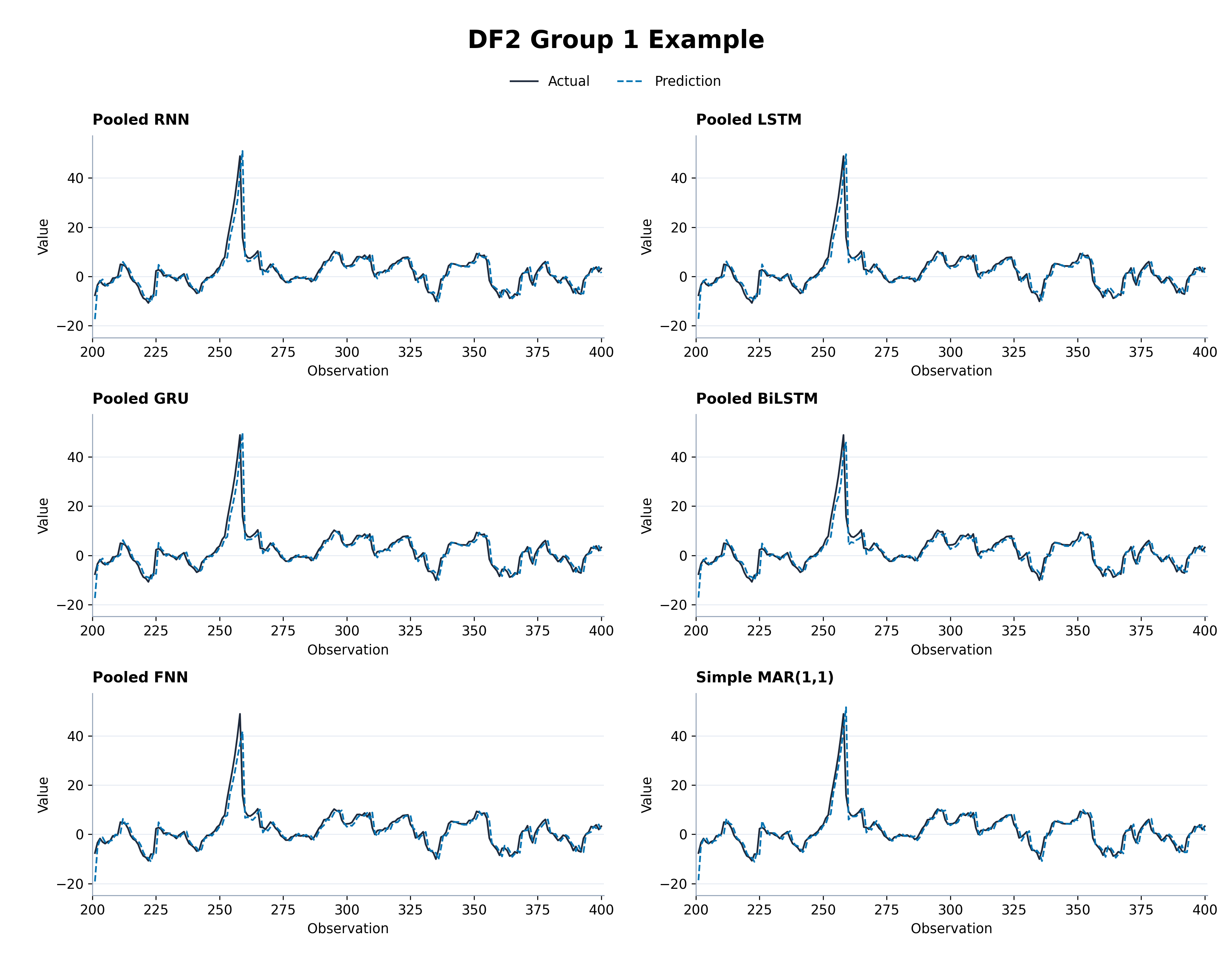}
        			\vspace{0.1cm}
			\hrule
			\vspace{0.1cm}
    \caption*{The panels show one-step-ahead forecast of the pooled mechanism for one simulated trajectory with degree of freedom 2 errors in Group 1.} 
    \label{pooled g1}
\end{figure}

The results from Group 1 in Figure \ref{pooled g1} show a remarkable improvement in bubble forecasting for all models.

\begin{figure}[H]
    \centering
    \caption{}
    \includegraphics[width=0.9\linewidth]{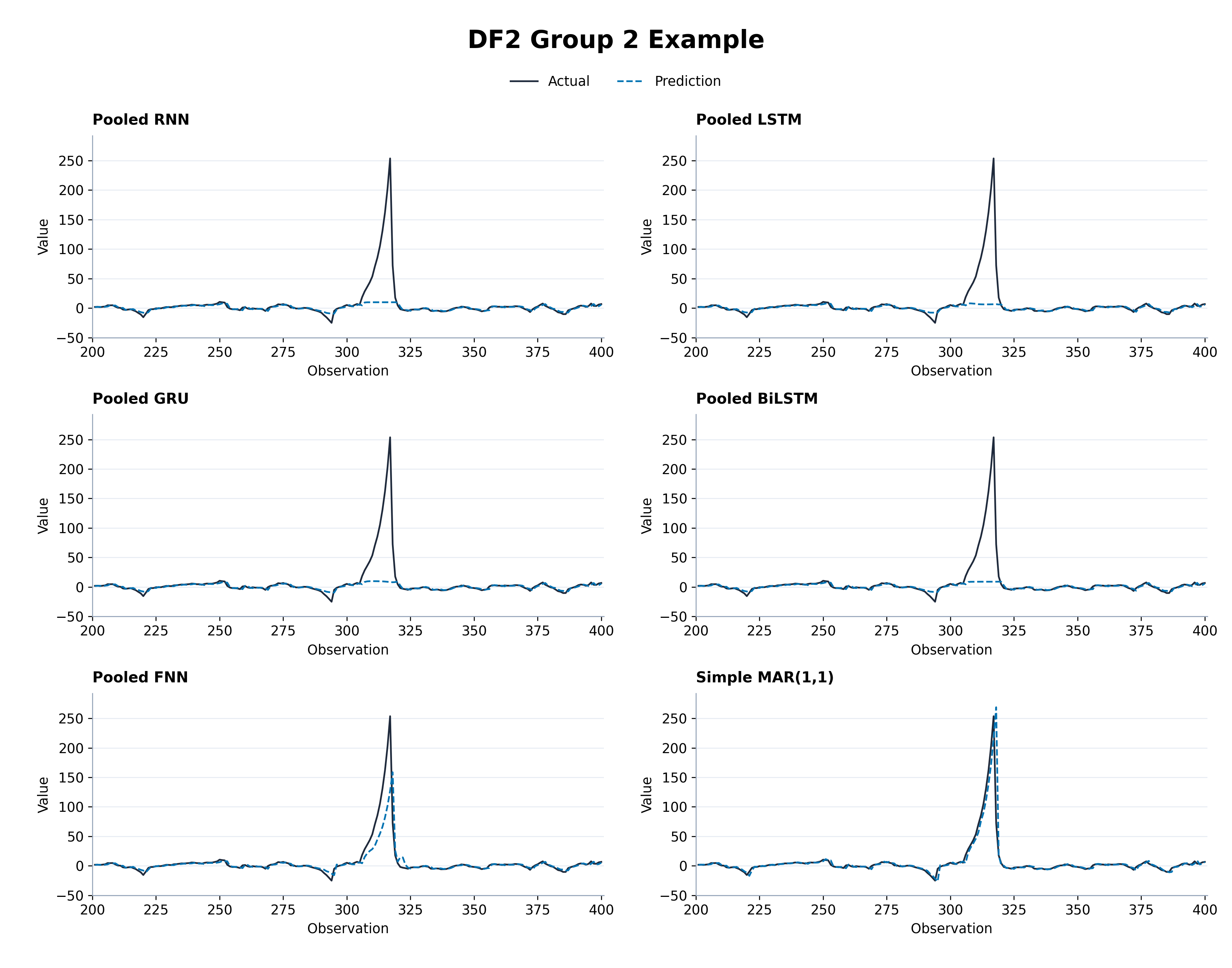}
        			\vspace{0.1cm}
			\hrule
			\vspace{0.1cm}
    \caption*{The panels show one-step-ahead forecast of the pooled mechanism for one simulated trajectory with degree of freedom 2 errors in Group 2.} 
    \label{pooled g2}
\end{figure}

Figure 11 illustrates that without experiencing the bubble scenario in the training set, increasing the size of the training set does not help the NN models forecast the bubble scenarios in the test sets. 

\begin{figure}
    \centering
    \caption{Pooled Test Sets Forecast RMSE for Bubble Episodes Distribution }
    \includegraphics[width=0.8\linewidth]{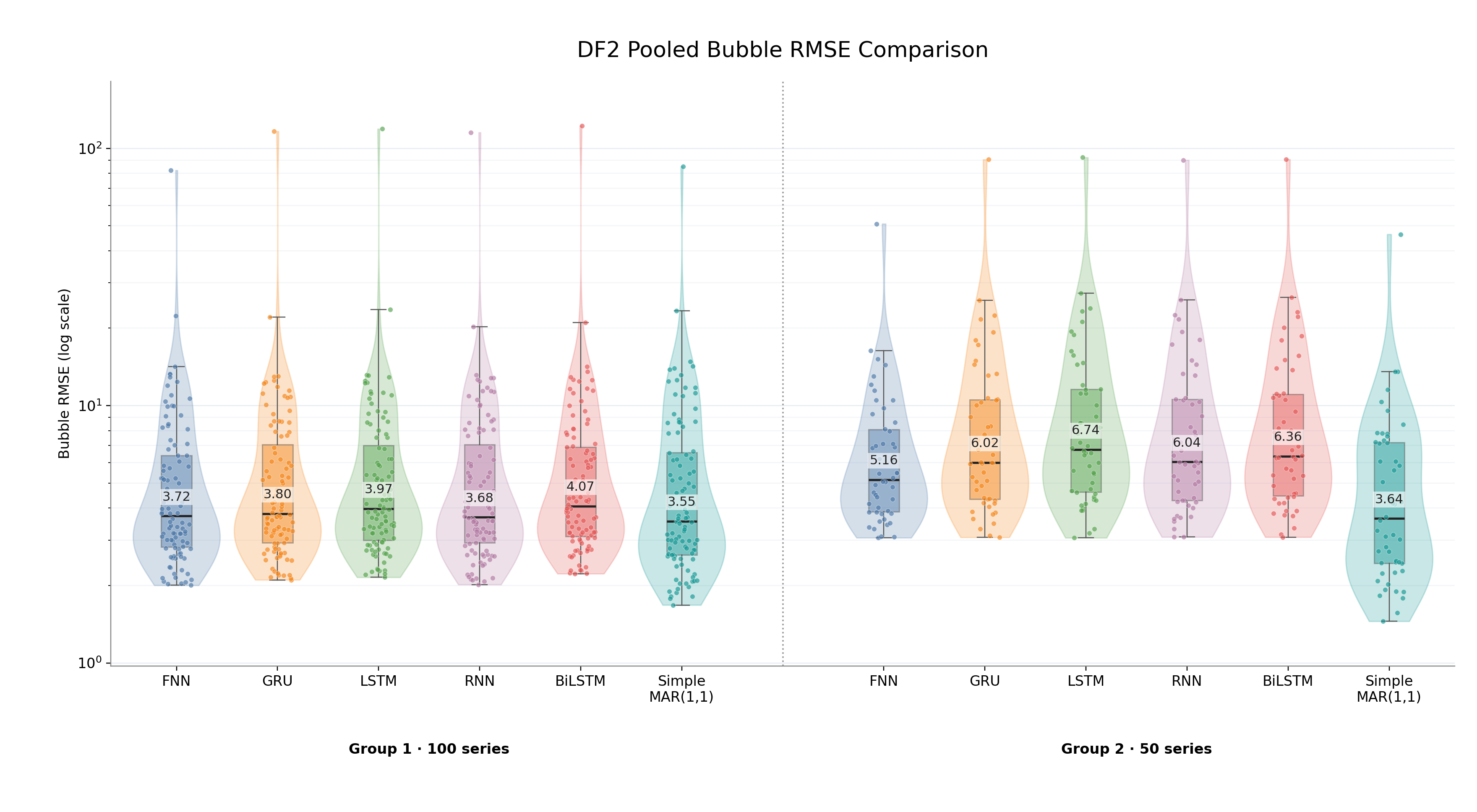}
    \caption*{The horizontal bars and values represent the median bubble RMSE for each group and algorithm.}
    \label{pooled bubble}
\end{figure}

Figure \ref{pooled bubble} below compares the RMSE of the nets computed from the extreme observations in the best-performing pooled mechanism with the simple forecast based on formula (2.4).

\nin We observe that the net algorithms show a relatively similar level of forecasting power. The RNN provides the lowest median RMSE for Group 1 (except the Simple MAR(1,1)). Meanwhile, FNN still shows an advantage in forecasting bubble episodes for Group 2 data.

\subsection{LSTM with Modified Predictive Criterion}

In this Section, we experiment with LSTM-based predictions and a modified training loss function to obtain better results. We replace the training MSE by loss functions that treat large errors differently, and experiment with likelihood-based loss functions. 

\subsubsection{MAE And Top-Percent MAE Losses}
 Top-percent MAE trains only on the largest absolute errors in each batch. For example, when the top 10 percent MAE is used as the fitting loss function, each batch update is driven by the largest 10 percent of absolute prediction errors.

\nin Settings used:
 Main early unit: 6;
 Epochs: 2,000;
 Batch size: 32;
 Optimizer: Adam;
 Learning rate: 0.001. 
 
We consider the following training loss functions: full-sample MAE, top 10 percent MAE, top 5 percent MAE, and top 1 percent MAE.  Let us introduce the following notation:
$$
    c_q = Q_{1-q/100}\!\left(|y_1|,\ldots,|y_N|\right),
\qquad
I_q = \{i : |y_i| > c_q\},
\qquad
n_q = |I_q|.
$$

\nin where \(c_q\) is the empirical \((100-q)\)th percentile of the
absolute observed values, \(I_q\) contains the indices exceeding
this threshold, and \(n_q\) is the number of selected observations.

The training loss functions: MAE, the top $q$ percent MAE and prediction error are defined as follows. A top-\(q\%\) MAE loss is

\begin{equation}
\mathcal{L}_{\mathrm{Top}\text{-}q\%\mathrm{\ MAE}}
=
\frac{1}{n_q}
\sum_{i \in I_q} |y_i-\hat{y}_i|,
\qquad n_q > 0.
\end{equation}

\subsubsection{t-Student Log Likelihood}
Next, we introduce a heavy-tail likelihood based on the t-Student distribution.
The t-Student  training loss function treats residuals as heavy-tailed random variables. The LSTM output is the conditional location, while degrees of freedom and scale parameters are learned.

\nin We use the following notation: MAR(1,1)-based (conditional) location: $\mu_t = \hat{y}_t$; residual: $e_t = y_t - \hat{y}_t$; degrees of freedom $\nu$: unknown and trainable; scale: unknown and trainable; constraint: $\nu > 1$ to ensure finite mean. The (negative) t-Student log-likelihood loss function is defined as: 
\begin{equation}
\mathcal{L}_{\mathrm{Student}\text{-}t}
=
\log \sigma
+\frac{1}{2}\log(\nu\pi)
+\log\Gamma\!\left(\frac{\nu}{2}\right)
-\log\Gamma\!\left(\frac{\nu+1}{2}\right)
+\frac{\nu+1}{2n}
\sum_{i=1}^{n}
\log\!\left(
1+\frac{(y_i-\hat{y}_i)^2}{\nu\sigma^2}
\right).
\end{equation}

\nin where \(n\) denotes the batch size. The degrees of freedom and scale are parameterized as:
\[
\nu = 1+\operatorname{softplus}(a)+\varepsilon,
\qquad
\sigma = \exp(s)+\varepsilon,
\qquad
\varepsilon = 10^{-6},
\]
where \(\operatorname{softplus}(a)=\log(1+\exp(a))\).
The parameters \(a\) and \(s\) are learned jointly with the
network parameters.

\subsubsection{t-Student Plus Top 10 Percent MAE}

Next we combine the t-Student likelihood with top 10 percent MAE. By doing this 
we combine heavy-tail residual modeling with an emphasis on the largest forecast errors.

\nin The training loss function is:
\[
\mathcal{L}=\mathcal{L}_{\mathrm{Student}\text{-}t}+\mathcal{L}_{Top10\%MAE}
\]

\nin where $\mathcal{L}_{\mathrm{Top}\text{-}q\%\mathrm{\ MAE}}$ is defined in (5.2) and $\mathcal{L}_{\mathrm{Student}\text{-}t}$ is defined in (5.3) 

\subsubsection{Regularized t-Student Plus Top 10 Percent MAE}

We add the L1, L2, and L1+L2 regularizations to the combined loss function considered above and use 16 and 32 units. The training loss function is:
\begin{equation}
\mathcal{L}=\mathcal{L}_{\mathrm{Student}\text{-}t}+\mathcal{L}_{Top10\%MAE}+\lambda_1\sum_j|\theta_j|+\lambda_2\sum_j\theta_j^2
\end{equation}

\nin where the last two terms represent the Lasso and Ridge regularizations, with $\lambda_1, \lambda_2$ obtained from cross-validation.

\begin{figure}[H]
\centering
\caption{Bubble Prediction Performance of Different Objective Functions}
\includegraphics[width=0.95\linewidth]{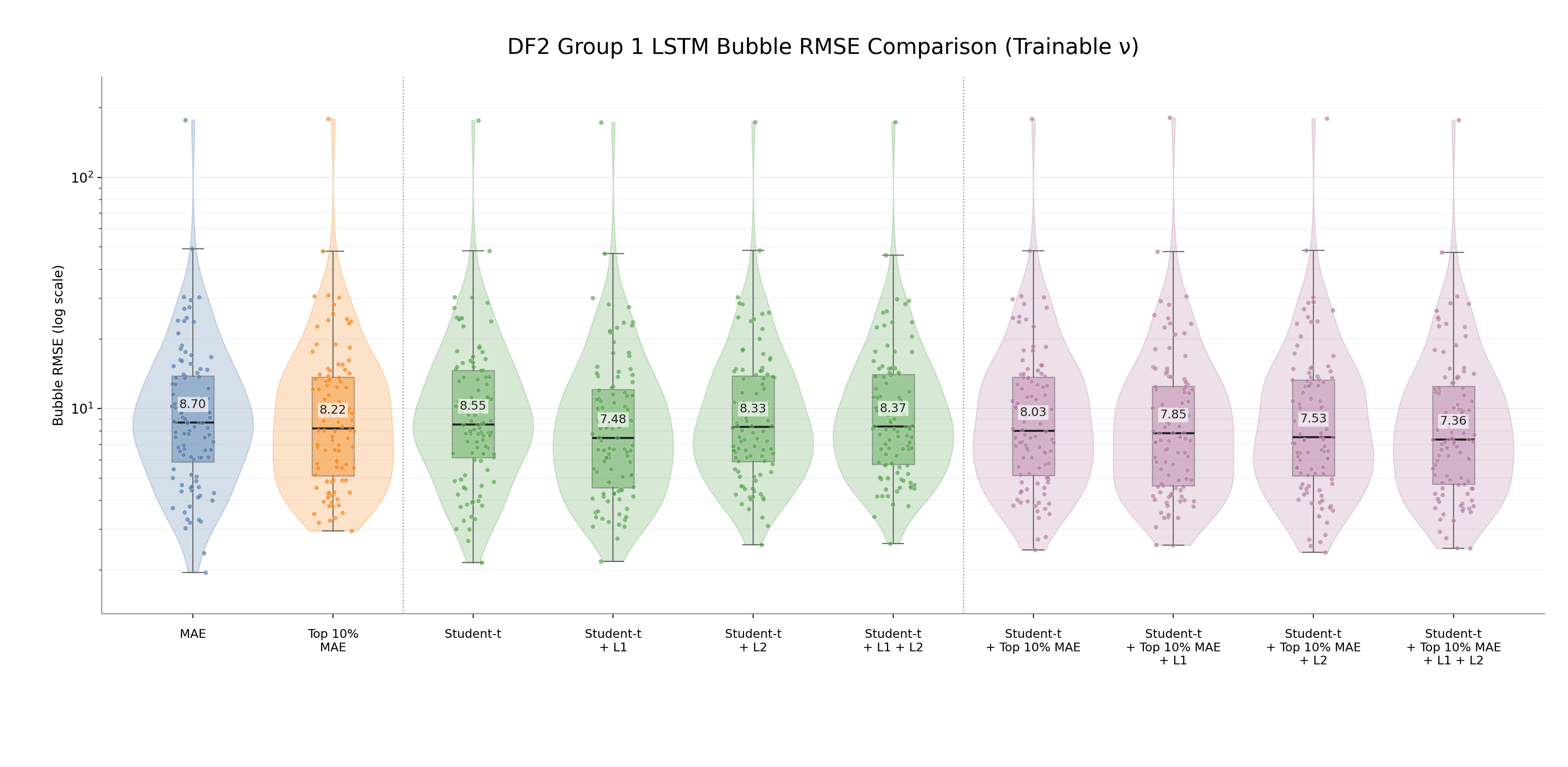}
\label{loss compare}
\end{figure}

Figure \ref{loss compare} compares the specifications with respect to the RMSE. Among the training loss functions evaluated, the Student-\(t\) negative log-likelihood combined with top \(10\%\) MAE and \(L_1+L_2\) regularization achieved the lowest median bubble-episode RMSE across the 100 degree of freedom 2 Group 1 series.

We conclude that 
 the training loss function based on combining the regularized Student-t likelihood and the top 10 percent MAE is our preferred method, providing the best forecast of the series.

\section{Conclusions}

We examined the performance of various neural nets in predicting extreme values observed during the bubble episodes that occur in MAR processes. We concluded that the presence of a bubble during the training period improves point predictions. The best predictions are obtained from the FNN. We pointed out that it involves the least amount of rescaling by activation functions and also preserves the Markov property of the MAR(1,1) process. Moreover, a pooling mechanism that increases the training set results in better predictions, although at the expense of increased computational complexity. In addition, we examined the LSTM to show that its predictive performance can be improved by changing the training criterion to better accommodate extreme values.

\newpage

\medskip

\nin \hrulefill

\section*{\normalsize Ethics Statement}
\addcontentsline{toc}{section}{Ethics Statement}

The authors have nothing to report.

\section*{\normalsize Conflicts of Interest}
\addcontentsline{toc}{section}{Conflicts of Interest}

The authors declare no conflicts of interest. 

\section*{\normalsize Data Availability Statement}
\addcontentsline{toc}{section}{Data Availability Statement}

The data used in this study were generated by simulation and are available from the corresponding author upon request.

\nocite{*}
\bibliographystyle{apalike}
\bibliography{references}

\newpage

\section*{Appendix 1}

\begin{table}[htbp]
\centering
\caption{Solo-Training Runtime for Cauchy Distribution Group 1 }
\label{tab:solo time}
\begin{tabular}{lc}
\hline
& Runtime (sec)\\
\hline
FNN & 6.58\\
RNN & 7.48\\
GRU & 7.02\\
LSTM & 6.92\\
LSTM--BiLSTM & 7.96\\
\hline
\end{tabular}
			
			\caption*{\small \textit {The runtimes in this table are recorded using the setting listed in Table \ref{hyper}. }}
\end{table}

\begin{table}[htbp]
\centering
\caption{Pooled Training Setting}
\label{tab:model_tuning_parameters_simple}
\begin{tabular}{lccccc}
\hline
& Units & Batch & Reg. & Epochs & Loss \\
\hline
FNN & 128 & 256 & 0.001 & 1000 & MSE \\
RNN & 256 & 128 & 0.001 & 1000 & MSE \\
GRU & 128 & 128 & 0.001 & 1000 & MSE \\
LSTM & 128 & 128 & 0.001 & 1000 & MSE \\
LSTM--BiLSTM & 64 & 128 & 0.001 & 1000 & MSE \\
\hline
\end{tabular}

\end{table}

\bigskip

\bigskip

\begin{table}[htbp]
\centering
\caption{Pooled-Training Runtime for Cauchy Distribution Group 1 }
\label{tab:pooled time}
\begin{tabular}{lc}
\hline
& Runtime (sec)\\
\hline
FNN & 6.45\\
RNN & 10.90\\
GRU & 7.92\\
LSTM & 7.04\\
LSTM--BiLSTM & 8.54\\
\hline
\end{tabular}
			
			\caption*{\small \textit {The runtimes in this table are recorded using the setting listed in Table \ref{tab:model_tuning_parameters_simple}. }}
\end{table}

\newpage 
\section*{Appendix 2}
\begin{center}
\textbf{Second-Order Expansion of the Asymptotic Variance}    
\end{center}

\nin This Appendix provides the steps of a proof of the second-order expansion of the  asymptotic variance:
$$
\text{Var}(\hat{\theta}_T) = S(T, \beta) = \sum_{n=1}^{T-1} \left[ n^{-2 \beta} \prod_{t=n+1}^T (1-t^{-\beta})^2 \right] + T^{-2 \beta}$$

\nin The asymptotic expansion of  \(S(T, \beta)\) when \(T \to \infty\), is as follows:

1. The Under-linear Regime (\(0 < \beta < 1\)). In this regime, the sum is dominated by the terms near the upper bound \(n \approx T\). The asymptotic expansion up to the second order is:

$$S(T, \beta) = \frac{1}{2} T^{-\beta} + \frac{1}{2} T^{-2\beta} + O\left(T^{-3\beta} + T^{-1-\beta}\right)$$

2. The Exact Case (\(\beta = 1\)). When \(\beta = 1\), the expression simplifies exactly for any \(T \ge 1\), leaving no higher-order terms:

$$S(T, 1) = \frac{1}{T}
$$

3. The Convergence Regime (\(\beta > 1\)). When \(\beta > 1\), the sequence converges to a strictly positive constant \(C(\beta)\) as \(T \to \infty\). The asymptotic expansion describes how the finite sum approaches this limit:

$$S(T, \beta) = C(\beta) - \frac{2}{\beta - 1} C(\beta) T^{1-\beta} + O\left(T^{-2(\beta-1)}\right)
$$

\nin where the constant \(C(\beta)\) is defined by the convergent infinite series:

$$C(\beta) = \sum_{n=1}^\infty \left[ n^{-2 \beta} \prod_{t=n+1}^\infty (1-t^{-\beta})^2 \right]
$$

\medskip

\begin{center}
\textbf{Derivation of the Second-Order Asymptotic Term for $0 < \beta < 1$}
\end{center}
The expression of $S(T, \beta)$ for $0 < \beta < 1$ is computed as follows. We consider the sum:
\[
S(T, \beta) = \sum_{n=1}^{T-1} \left[ n^{-2 \beta} \prod_{t=n+1}^T (1-t^{-\beta})^2 \right] + T^{-2 \beta}
\]
To find the second-order term when $T \to \infty$ and $0 < \beta < 1$, we analyze the behavior of the internal product and convert the sum into an integral near the upper bound $T$.

\nin {\it Step 1: The Internal Product} \\
Let $P_n(T) = \prod_{t=n+1}^T (1-t^{-\beta})^2$. By taking the natural logarithm, we can rewrite the product as a sum:
\[
\ln P_n(T) = 2 \sum_{t=n+1}^T \ln(1-t^{-\beta})
\]
Using the Taylor expansion of $\ln(1-x) = -x - \frac{x^2}{2} + O(x^3)$ for large $t$, we obtain:
\[
\ln P_n(T) = 2 \sum_{t=n+1}^T \left( -t^{-\beta} - \frac{1}{2}t^{-2\beta} + O(t^{-3\beta}) \right)
\]
Applying the integral approximation (Euler-Maclaurin formula) for the sums of powers yields:
\[
\sum_{t=n+1}^T t^{-\beta} = \frac{T^{1-\beta} - n^{1-\beta}}{1-\beta} + O(T^{-\beta})
\]
Exponentiating back, the product behaves as:
\[
P_n(T) \approx \exp\left( -2\frac{T^{1-\beta} - n^{1-\beta}}{1-\beta} \right) \left[ 1 - \frac{T^{1-2\beta} - n^{1-2\beta}}{1-2\beta} + \dots \right]
\]

\nin {\it Step 2: Change of Variables and Local Expansion} \\
The main contribution to the sum $S(T, \beta)$ comes from the terms where $n$ is close to $T$. We introduce the change of variable $n = T - x$, where $x$ represents the distance from the upper bound $T$, and expand in terms of $x/T$:
\begin{enumerate}
    \item The prefactor $n^{-2\beta}$:
    \[
    (T-x)^{-2\beta} = T^{-2\beta} \left(1 - \frac{x}{T}\right)^{-2\beta} = T^{-2\beta} \left( 1 + 2\beta\frac{x}{T} + O\left(\frac{x^2}{T^2}\right) \right)
    \]
    \item The exponential argument:
    
    By Taylor expansion of the primitive difference near $x=0$:
    \[
    \frac{2}{1-\beta} \left( (T-x)^{1-\beta} - T^{1-\beta} \right) = -2T^{-\beta}x - \beta T^{-\beta-1}x^2 + O(T^{-\beta-2}x^3)
    \]
\end{enumerate}

\nin {\it Step 3: Integral Evaluation of the Terms}\\
We substitute these expansions into the sum and approximate it as a continuous integral over $x$ from $0$ to $\infty$ to obtain the following integrand:
\[
S(T, \beta) \approx \int_{0}^{\infty} T^{-2\beta} \left( 1 + \frac{2\beta x}{T} \right) e^{-2T^{-\beta}x} \left( 1 - \beta T^{-\beta-1}x^2 \right) \mathrm{d}x
\]
Expanding this expression produces three main integrals:
\begin{enumerate}
    \item First-order (dominant) term:
    
    \[
    \int_{0}^{\infty} T^{-2\beta} e^{-2T^{-\beta}x} \mathrm{d}x = T^{-2\beta} \left[ \frac{e^{-2T^{-\beta}x}}{-2T^{-\beta}} \right]_{0}^{\infty} = \frac{1}{2} T^{-\beta}
    \]
    \item Second-order correction from the prefactor:
    
    We use $\int_{0}^{\infty} x e^{-ax} \mathrm{d}x = \frac{1}{a^2}$ with $a = 2T^{-\beta}$:
    \[
    \int_{0}^{\infty} T^{-2\beta} \left(\frac{2\beta x}{T}\right) e^{-2T^{-\beta}x} \mathrm{d}x = \frac{2\beta}{T^{1+2\beta}} \frac{1}{4T^{-2\beta}} = \frac{\beta}{2} T^{-1}
    \]
    \item Second-order correction from the exponential expansion:
    
    We use $\int_{0}^{\infty} x^2 e^{-ax} \mathrm{d}x = \frac{2}{a^3}$:
    \[
    \int_{0}^{\infty} T^{-2\beta} \left(-\frac{\beta x^2}{T{\beta+1}}\right) e^{-2T^{-\beta}x} \mathrm{d}x = -\frac{\beta}{T^{3\beta+1}} \frac{2}{8T^{-3\beta}} = -\frac{\beta}{4} T^{-1}
    \]
\end{enumerate}

\nin {\it Step 4: Discrete Corrections}\\
We sum up the continuous corrections to get $\frac{\beta}{2}T^{-1} - \frac{\beta}{4}T^{-1} = \frac{\beta}{4}T^{-1}$. Since $0 < \beta < 1$, this term is of higher order compared to the discrete boundary correction. 

By applying the precise discrete-to-continuous correction from the Euler-Maclaurin formula, the boundary term at $n=T$ provides a direct $\frac{1}{2} T^{-2\beta}$ contribution (which matches the only $T^{-2\beta}$ boundary term in the original formulation). By combining the dominant term and this boundary correction we get the final second-order asymptotic expansion:
\[
S(T, \beta) = \frac{1}{2} T^{-\beta} + \frac{1}{2} T^{-2\beta} + O\left(T^{-3\beta} + T^{-1-\beta}\right)
\]

\end{document}